\documentclass[twocolumn]{aastex63}
\usepackage[utf8]{inputenc} 
\usepackage[T1]{fontenc}    
\usepackage{hyperref}       
\usepackage{url}            
\usepackage{booktabs}       
\usepackage{amsfonts}       
\usepackage{nicefrac}       
\usepackage{microtype}      
\usepackage{amsmath}
\usepackage{ragged2e}
\usepackage{float}
\usepackage{graphicx}
\usepackage{natbib}
\usepackage[english]{babel}
\usepackage[normalem]{ulem}
\usepackage{cancel}
\usepackage{lineno}

\DeclareMathOperator\erf{erf}

\graphicspath{{./}{figures/}}
\begin{document}

\title{The Kinetic Expansion of Solar-Wind Electrons: Transport Theory and Predictions for the very Inner Heliosphere}

\author{Seong-Yeop Jeong}
\affiliation{Mullard Space Science Laboratory, University College London, Dorking, RH5 6NT, UK; s.jeong.17@ucl.ac.uk}

\author{Daniel Verscharen}
\affiliation{Mullard Space Science Laboratory, University College London, Dorking, RH5 6NT, UK; s.jeong.17@ucl.ac.uk}

\author{Christian Vocks}
\affiliation{Leibniz-Institut für Astrophysik Potsdam (AIP), An der Sternwarte 16, D-14482 Potsdam, Germany}

\author{Joel B. Abraham}
\affiliation{Mullard Space Science Laboratory, University College London, Dorking, RH5 6NT, UK; s.jeong.17@ucl.ac.uk}

\author{Christopher J. Owen}
\affiliation{Mullard Space Science Laboratory, University College London, Dorking, RH5 6NT, UK; s.jeong.17@ucl.ac.uk}

\author{Robert T. Wicks}
\affiliation{Northumbria University, Newcastle, NE1 8ST, UK}

\author{Andrew N. Fazakerley}
\affiliation{Mullard Space Science Laboratory, University College London, Dorking, RH5 6NT, UK; s.jeong.17@ucl.ac.uk}

\author{David Stansby}\affiliation{Mullard Space Science Laboratory, University College London, Dorking, RH5 6NT, UK; s.jeong.17@ucl.ac.uk}

\author{Laura Ber\v{c}i\v{c}}
\affiliation{Mullard Space Science Laboratory, University College London, Dorking, RH5 6NT, UK; s.jeong.17@ucl.ac.uk}

\author{Georgios Nicolaou}
\affiliation{Southwest Research Institute, San Antonio, TX 78238, USA}

\author{Jeffersson A. Agudelo Rueda}
\affiliation{Mullard Space Science Laboratory, University College London, Dorking, RH5 6NT, UK; s.jeong.17@ucl.ac.uk}

\author{Mayur Bakrania}
\affiliation{Mullard Space Science Laboratory, University College London, Dorking, RH5 6NT, UK; s.jeong.17@ucl.ac.uk}


\begin{abstract}
We propose a transport theory for the kinetic evolution of solar-wind electrons in the heliosphere. We derive a gyro-averaged kinetic transport equation that accounts for the spherical expansion of the solar wind and the geometry of the Parker-spiral magnetic field. To solve our three-dimensional kinetic equation, we develop a mathematical approach that combines the Crank--Nicolson scheme in velocity space and a finite-difference Euler scheme in configuration space. We initialize our model with isotropic electron distribution functions and calculate the kinetic expansion at heliocentric distances from 5 to 20 solar radii. In our kinetic model, the electrons evolve mainly through the combination of the ballistic particle streaming, the magnetic mirror force, and the electric field. By applying fits to our numerical results, we quantify the parameters of the electron strahl and core part of the electron velocity distributions. The strahl fit parameters show that the density of the electron strahl is around 7\% of the total electron density at a distance of 20 solar radii, the strahl bulk velocity and strahl temperature parallel to the background magnetic field stay approximately constant beyond a distance of 15 solar radii, and $\beta_{\parallel s}$ (i.e., the ratio between strahl parallel thermal pressure to the magnetic pressure) is approximately constant with heliocentric distance at a value of about 0.02. We compare our results with data measured by Parker Solar Probe. Furthermore, we provide theoretical evidence that the electron strahl is not scattered by the oblique fast-magnetosonic/whistler instability in the near-Sun environment. \\
\\
\textit{Unified Astronomy Thesaurus concepts}: Solar wind (1534); Space plasma (1544); Heliosphere (711); Theoretical model (2107)
\end{abstract}



\section{Introduction} \label{intro}
The solar-wind plasma consists of positively charged ions and negatively charged electrons. The solar-wind electrons play important roles for the evolution of the solar wind. They guarantee the overall plasma quasi-neutrality and  provide significant  heat flux through  non-thermal properties of the electron velocity distribution functions  \citep[VDFs;][]{2006LRSP....3....1M}. Moreover, the electrons generate a global ambipolar electric field through their thermal pressure gradient \citep{1970A&A.....6..219J}.  

\textit{In-situ} measurements of the electron VDF in the solar wind reveal multiple deviations from a Maxwellian equilibrium state \citep{doi:10.1029/JA092iA02p01075,doi:10.1029/2008JA013883}. The electron VDF typically consists of three different electron populations: the core, the halo, and the strahl. The electron core, which accounts for most of the electrons in the solar wind, has a relatively low energy ($\lesssim 50\,\mathrm{eV}$) and is nearly isotropic. The electron halo has a higher energy ($\gtrsim50\,\mathrm{eV}$) than the core and is nearly isotropic as well. Lastly, the electron strahl is an energetic and highly field-aligned electron population, and carries most of the heat flux \citep{https://doi.org/10.1029/JA080i031p04181,https://doi.org/10.1029/JA092iA02p01103}. However, the formation and scattering mechanisms of the electron strahl in the heliosphere are still unclear.

In order to model the evolution of solar-wind electrons and non-Maxwellian features in the electron VDF, a kinetic approach is necessary. Previous theoretical models for the evolution of the electron VDF primarily account for the global temperature gradient, magnetic mirror forces, and wave--particle interactions near the Sun \citep{1997JGR...102.4701L,2000JGR...105...35L,Vocks_2003,https://doi.org/10.1029/2008JA013294,Smith_2012,Landi_2012,Seough_2015,Tang_2020,2020}. Coulomb collisions affect the evolution of the electron VDF in the solar wind near the corona, which has important implications for exospheric solar-wind models \citep{1970A&A.....6..219J,refId0,Zouganelis_2005}. 
However, at large distances from the Sun, other mechanisms must be considered for local strahl scattering \citep{2018MNRAS.474..115H,2019MNRAS.484.2474H,2019MNRAS.489.3412B}. For instance, the strahl-driven oblique fast-magnetosonic/whistler (FM/W) instability has recently received much attention as such a mechanism  \citep{2019ApJ...871L..29V,2019ApJ...886..136V,L_pez_2020,Jeong_2020,Micera_2020,Micera_2021,refId00,Sun_2021}.

Since the gradients in the plasma and field parameters (e.g., gradients of the solar-wind speed, temperature, and magnetic field) are greater at smaller heliocentric distances, we expect that the electron VDF undergoes a stronger modification near the Sun. Therefore, it is important to model the electron VDF evolution near the Sun, especially in regions that we have not explored with spacecraft yet. In the acceleration region of the solar wind \citep{Bemporad_2017,YAKOVLEV2018552}, the electron number density profile exhibits a steeper  decrease  than a $1/r^2$-profile. In that region, the pressure gradient has a significant impact on ballistic particle streaming along the magnetic field and the creation of an ambipolar electric field. This ambipolar electric field returns a large number of the streaming electrons back to the Sun \citep{Boldyrev9232}. In addition, the magnetic mirror force in the decreasing magnetic field focuses outward-streaming electrons towards narrow pitch-angles. However, the magnetic mirror force becomes ineffective at large distances from the Sun due to the weakened gradient of the magnetic field \citep{https://doi.org/10.1029/2008JA013294}. At the same time, Coulomb collisions occur more frequently at small heliocentric distances \citep{https://doi.org/10.1029/JA091iA07p08045}. 

For the understanding of local strahl scattering, it is  important to model the evolution of the electron VDF  up to large heliocentric distances where the effect of the Parker-spiral geometry of the interplanetary magnetic field  is noticeable \citep{2018MNRAS.480.1499H,2021MNRAS.507.1329S,refId00}. Many previous studies for the radial kinetic evolution are based on a simplified radial magnetic-field geometry. However, exospheric models suggest that the inclusion of a more realistic, non-radial  magnetic field significantly modifies the kinetic  properties of the expanding plasma at heliocentric distances beyond $100 r_s$ \citep{https://doi.org/10.1029/JA077i001p00001,https://doi.org/10.1029/2000GL011888}, where $r_s$ is the solar radius.  Moreover, for a comparison of the electron VDF with observations, analytical models must quantify both the bulk parameters and the shape of the electron VDF.

In Section~\ref{2}, we derive a gyro-averaged kinetic transport equation that accounts for the spherical expansion of the solar wind and the geometry of the Parker-spiral magnetic field in the heliosphere. Our derivation leads to a  kinetic transport equation consistent with the transport equations derived by \citet{1971ApJ...170..265S}, \citet{1985ApJ...296..319W}, \citet{https://doi.org/10.1029/96JA03671}, \citet{2007ApJ...662..350L}, \citet{le_Roux_2009} and \citet{book}. 
In Section~\ref{3}, we lay out our numerical treatment for our kinetic transport equation. In Section~\ref{4}, we model the kinetic expansion of solar-wind electrons from the corona at a heliocentric distance of $5r_s$, where collisions are more important, to a heliocentric distance of $20r_s$. By applying a fitting scheme to our modeled electron VDFs, we analyze the evolution of the fit parameters with heliocentric distance in the spherically expanding solar wind. We then compare our fit parameters with measurements from Parker Solar Probe (PSP). Furthermore, we show that, at heliocentric distances below $20r_s$, the generated electron strahl is not scattered by the oblique FM/W instability. In Section~\ref{5}, we discuss and conclude our results. In Appendix~\ref{A}, we present our mathematical strategy for the solution of our three-dimensional kinetic transport equation (2D in velocity space and 1D in configuration space) based on the combination of a Crank--Nicolson scheme in velocity space and a finite-difference Euler scheme in configuration space. In Appendix~\ref{b}, we discuss the effect of our numerical smoothing algorithm in velocity space. 


\section{Kinetic Transport Theory} \label{2}
In this section, we  derive a gyro-averaged kinetic transport equation accounting for the non-radial, average spiral shape of the interplanetary magnetic field. 
Our kinetic transport equation  describes the radial evolution of the electron VDF in the spherically expanding solar wind.

\subsection{Non-Inertial Co-moving Reference Frame}\label{2.1}
Due to the Sun's rotation, the heliospheric magnetic field follows on average the Parker spiral \citep{1958ApJ...128..664P}. The spiral structure begins radially near the Sun and then exhibits an increasing relative contribution from the magnetic field's azimuthal component.
\begin{figure}[ht]
  \centering
  \includegraphics[width=\columnwidth]{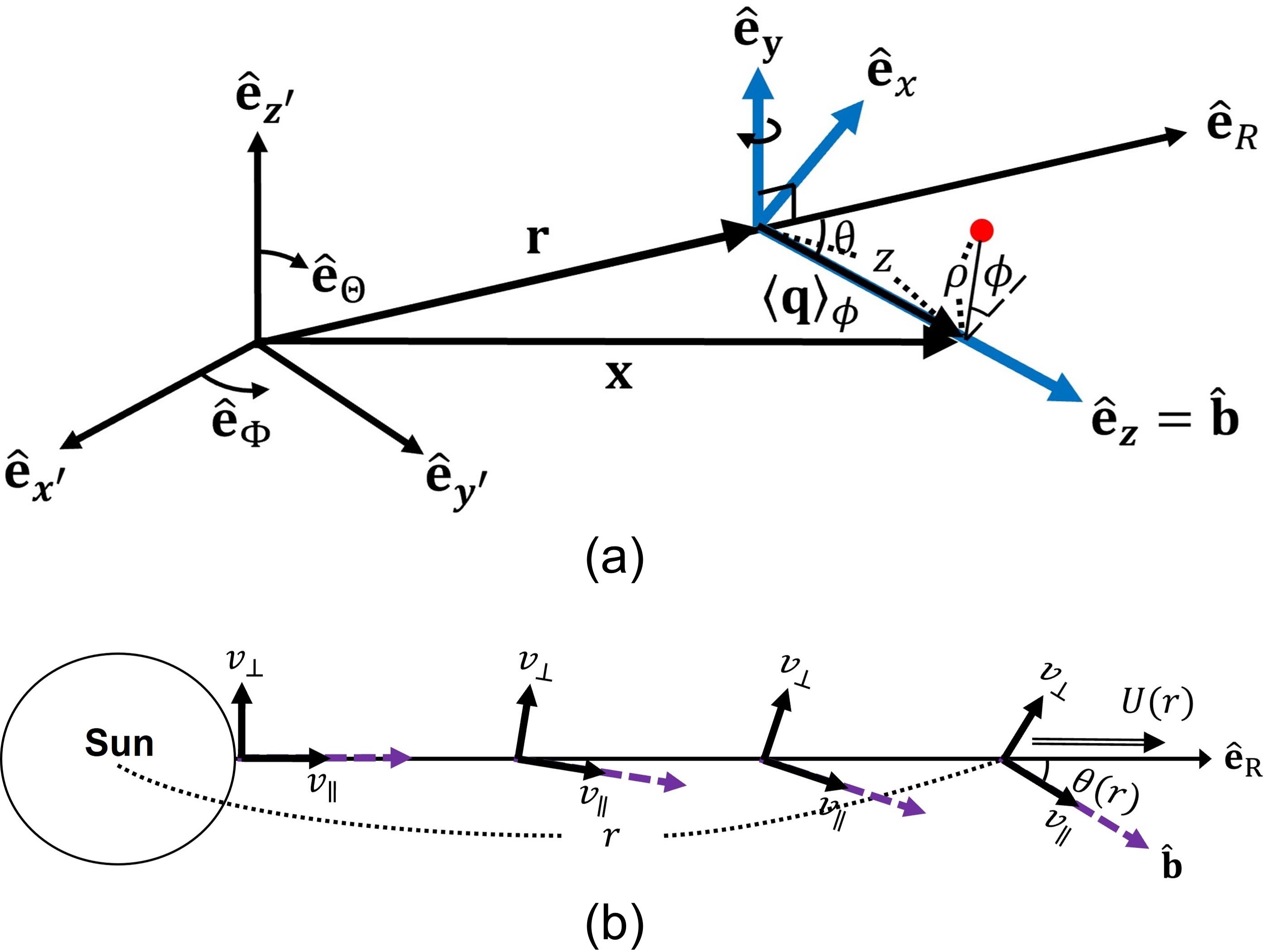}
  \caption{(a) Schematic of our reference frames. The Sun-at-rest frame (black axes) is described by the spherical coordinates $\mathbf x=(R, \Theta,\Phi)$, and the co-moving wind frame (blue axes) is described by the cylindrical coordinates $\mathbf q= (\rho, \phi, z)$.  The $z$-axis of the co-moving wind frame is parallel to the direction of the local magnetic field at distance $r$, and $\theta$ is the angle between the $z$-axis and the $R$-axis. (b) Evolution of the velocity-space coordinates in the co-moving reference frame depending on distance from the Sun. The purple dashed arrow indicates the local magnetic field.}
  \label{Fig1}
\end{figure}
In Fig.~\ref{Fig1}a, we define two reference frames: (i) the Sun-at-rest frame in spherical coordinates $\mathbf x=(R, \Theta,\Phi)$ and (ii) the co-moving wind frame (blue axes) in cylindrical coordinates $\mathbf q= (\rho, \phi, z)$. The vector $\mathbf r=\hat {\mathbf e}_R r$ describes the origin of the co-moving wind frame in the Sun-at-rest frame. 
The origin of the co-moving wind frame moves into the anti-sunward direction in the Sun-at-rest frame with the wind speed $\mathbf{U}(\mathbf x)=\dot {\mathbf r}= \hat{\mathbf e}_R U$ along the radial direction. We set our solar-wind speed profile $\mathbf{U}$ and our background magnetic field  profile $\mathbf{B}(\mathbf x)=\hat{\mathbf e}_R B_R- \hat{\mathbf e}_\Phi B_\Phi$ so that $\nabla \times (\mathbf U \times \mathbf{B})=0$. We require that the $z$-axis of the co-moving wind frame is always parallel to the direction of the local magnetic field at distance $r$. Our wind frame is a non-inertial reference frame due to the acceleration of $\mathbf U$ and the rotation of the reference frame from the $r$-dependence of the angle $\theta$ between $\hat{\mathbf e}_R$ and $\hat{\mathbf b}$, so that fictitious forces occur. 

In Fig.~\ref{Fig1}a, the red circle represents a test particle. We define the vector $\mathbf x^{\prime}$ as the spatial coordinate of the test particle:
\begin{equation}
\mathbf x^{\prime} = \mathbf r + \mathbf q
\end{equation}
and the vector $\mathbf x$ as its gyro-averaged guiding center:
\begin{equation}\label{rs}
\begin{split}
\mathbf x &= \mathbf r +\left \langle \mathbf q \right \rangle_\phi=\mathbf r+ \hat{\mathbf b}z,
\end{split}
\end{equation}
where 
\begin{equation}\label{spatialgyroaverage}
\begin{split}
\langle \alpha \rangle_\phi \equiv \frac{1}{2\pi}\int_{0}^{2\pi} \alpha \,d{\phi},
\end{split}
\end{equation}
and $\alpha$ is an arbitrary function of $\phi$. Using Parker's model \citep{1958ApJ...128..664P}, we define the unit vector along $\mathbf B$ as 
\begin{equation}
\hat {\mathbf b}=\frac{\mathbf B}{|\mathbf B|}=\hat{\mathbf e}_R \frac{B_R}{B_z} - \hat{\mathbf e}_\Phi \frac{B_\Phi}{B_z} =\hat{\mathbf e}_R \cos\theta- \hat{\mathbf e}_\Phi \sin\theta 
\end{equation}
where
\begin{equation}\label{Bz}
B_z(r)=\sqrt{B_R^2(r)+B_\Phi^2(r)},
\end{equation}
\begin{equation}\label{Br}
B_R(r)=B_0 \left(\frac{r_0}{r} \right)^2,
\end{equation}
\begin{equation}\label{Bp}
B_\Phi(r)=B_0\frac{\Omega_\odot}{U(r)}\frac{r_0^2}{r} \sin\Theta,
\end{equation}
$B_0$ is the reference value of the radial component of the magnetic field at the reference distance $r_0$, $\Omega_\odot$ is the Sun's rotation frequency and $\Theta$ is constant.

Fig.~\ref{Fig1}b illustrates the radial evolution of the velocity-space coordinates in the co-moving solar-wind frame, rotating in accordance with the Parker-spiral geometry. For any given particle, we define its velocity coordinates in the directions perpendicular and parallel with respect to the background magnetic field in the co-moving solar-wind frame as 
\begin{equation}\label{vper}
v_\perp=\sqrt{\dot \rho^2+\rho^2 \dot \phi^2}
\end{equation}
and
\begin{equation}\label{vpal}
v_\parallel=\dot z,
\end{equation}
so that  $\mathbf v= \dot{  \mathbf q } =\hat{\mathbf e}_x v_\perp\cos\phi_v+\hat{\mathbf e}_y v_\perp\sin\phi_v+ \hat{\mathbf b} v_\parallel$, where $\phi_v$ is the azimuthal angle of the velocity vector in the cylindrical co-moving reference frame. We note that $\hat{\mathbf e}_\perp=\hat{\mathbf e}_x \cos\phi_v + \hat{\mathbf e}_y \sin\phi_v$.

\subsection{Kinetic Expansion of Solar-Wind Electrons}\label{2.2}
To study the kinetic evolution of solar-wind electrons along the radial direction, as shown in Fig.~\ref{Fig1}b, we define the electron VDF in six-dimensional phase-space and time as
\begin{equation}\label{f}
f_e \equiv f_e\left( \mathbf x^{\prime}, \mathbf v, t \right).
\end{equation}
We define the coordinates so that the configuration space coordinates $\mathbf x^{\prime}$ are in the Sun-at-rest frame while the velocity space coordinates $\mathbf v$ are in the co-moving wind frame. The subscript $e$ indicates electron quantities.  
Based on Eq.~(\ref{f}), we calculate the kinetic evolution of $f_e$ under the action of  ballistic particle streaming, internal and external forces, and Coulomb collisions. In the present paper, we do not include a term for local wave--particle interactions in our equation. We evaluate the total time derivative of Eq.~(\ref{f}) along particle trajectories in phase space according to Liouville's theorem as
\begin{equation}\label{fulconvec}
\begin{split}
\frac{\partial f_e}{\partial t}\!+\!(\mathbf U+\mathbf v )\!\cdot \!\nabla_{\mathbf x^{\prime}} f_e&\!-\!\Big\{\!\big[(\mathbf U+\mathbf v )\!\cdot\! \nabla_{\mathbf x^{\prime}}\big] \mathbf U \!\Big\} \!\cdot\! \nabla_{\mathbf v} f_e\\
&+\dot{\mathbf v}\cdot \nabla_{\mathbf v} f_e=\left(\frac{\partial f_e }{\partial t}\right)_{\mathrm{col}},
\end{split}
\end{equation}
where
\begin{equation}\label{gradv}
\begin{split}
\nabla_{\mathbf v} \equiv \hat{\mathbf b} \frac{\partial }{\partial v_\parallel}+\hat{\mathbf e}_\perp \frac{\partial }{\partial v_\perp}+\hat{\mathbf e}_{\phi_v} \frac{1}{v_\perp}\frac{\partial }{\partial \phi_v}.
\end{split}
\end{equation}
We discuss the Coulomb-collision term $(\partial f_e/\partial t)_{\mathrm{col}}$ in Section~\ref{2.5}. On the left-hand side of Eq.~(\ref{fulconvec}), the first term describes the explicit variation of $f_e$ with time, the second term quantifies the ballistic particle streaming, the third term corresponds to internal forces caused by the velocity transformation into the (accelerating) co-moving wind frame, and the fourth term corresponds to external forces. We assume that these external forces are only due to the electromagnetic field.  

By assuming that the electron's gyro-period is much smaller than the other involved time scales, $f_e$ can be safely assumed to be gyrotropic and a function of $\mathbf x$ instead of $\mathbf x^{\prime}$, so that $f_e\equiv f_e(\mathbf x,v_{\perp},v_{\parallel},t)$. We then apply gyro-phase averaging to Eq.~(\ref{fulconvec}) as
\begin{equation}\label{gyrofulconvec}
\begin{split}
&\frac{\partial f_e}{\partial t}\!+\!\left(\mathbf U\!+\!\langle \mathbf v \rangle_{\!\phi_v} \right)\!\cdot\! \nabla_{\mathbf x} f_e\!-\!\Big\langle\!\big[ (\mathbf v \! \cdot \!\nabla_{\mathbf x}) \mathbf U \big] \! \cdot \! \hat{\mathbf e}_\perp \!\Big\rangle_{\!\phi_v} \! \frac{\partial f_e}{\partial v_\perp}\\
&-\!\Big\{\big[\left(\mathbf U\!+\!\langle \mathbf v \rangle_{\!\phi_v} \right)\!\cdot\! \nabla_{\mathbf x}\big] \mathbf U \Big\}\! \cdot \! \hat{\mathbf b} \frac{\partial f_e}{\partial v_\parallel}\\
&+\!\left\langle \dot{v}_\parallel \right\rangle_{\!\phi_v} \!\!\frac{\partial f_e}{\partial v_\parallel}\!+\!\left\langle\dot{ v}_\perp\right\rangle_{\!\phi_v}\!\! \frac{\partial f_e}{\partial v_\perp}
=\!\left(\frac{\partial f_e }{\partial t}\right)_{\mathrm{col}},
\end{split}
\end{equation}
where
\begin{equation}\label{gyroaverage}
\begin{split}
\langle \beta \rangle_{\phi_v} \equiv \frac{1}{2\pi}\int_{0}^{2\pi} \beta d{\phi_v},
\end{split}
\end{equation}
and $\beta$ is an arbitrary function of $\phi_v$. 


\subsection{Hamiltonian Analysis of External Forces}\label{2.3}
We apply the Hamiltonian formalism to analyze the external forces exerted on the electrons in the co-moving wind frame, corresponding to the last two terms on the left-hand side of Eq.~(\ref{gyrofulconvec}). The fundamental mathematics is explained in great detail by \citet{Nolting2016} and \citet{gurnett_bhattacharjee_2017}. 

We begin our analysis by defining the Lagrangian $\mathcal L$ in the generalized cylindrical  coordinates $\mathbf q= (\rho, \phi, z)$ in the co-moving wind frame as
\begin{equation}\label{lag}
\begin{split}
\mathcal L=\frac{1}{2}m_e  \mathbf v^2 +\frac{q_e}{c} \left( \mathbf A \cdot \mathbf v \right)-q_e\varphi,
\end{split}
\end{equation}
where $\mathbf v= \hat {\mathbf e}_\rho \dot\rho  + \hat {\mathbf e}_\phi \rho \dot \phi + \hat {\mathbf b} v_\parallel $, $q_e$ and $m_e$ are the charge and mass of an electron ($q_e=-e$), and $c$ is the speed of light. We assume that, on average, the electric and magnetic fields are static and only depend on the configuration space coordinate $\mathbf x^{\prime}$ so that
 $\mathbf{E}(\mathbf x^{\prime})= \hat {\mathbf e}_\rho E_\rho+ \hat {\mathbf e}_\phi E_\phi+ \hat {\mathbf b} E_z$ and $\mathbf{B}(\mathbf x^{\prime})= \hat {\mathbf e}_\rho B_\rho+ \hat {\mathbf e}_\phi B_\phi+ \hat {\mathbf b} B_z$. We define the scalar potential $\varphi(\mathbf x^{\prime})$ so that $\mathbf E=-\nabla_{\mathbf q} \varphi$, and the vector potential $\mathbf{A}(\mathbf x^{\prime})= \hat {\mathbf e}_\rho A_\rho+ \hat {\mathbf e}_\phi A_\phi+ \hat {\mathbf b} A_z$ so that $\mathbf B=\nabla_{\mathbf q} \times \mathbf A$.  The components of the generalized momentum $\mathbf p$ are then given as
\begin{equation}\label{genemorho}
p_\rho=m_e \dot \rho + \frac{q_e}{c} A_\rho,
\end{equation}
\begin{equation}\label{genemophi}
p_\phi=m_e \rho^2 \dot \phi + \frac{q_e}{c} \rho A_\phi,
\end{equation}
and
\begin{equation}\label{genemoz}
p_z=m_e v_\parallel + \frac{q_e}{c} A_z.
\end{equation}
Assuming that the scalar potential $\varphi$  depends only on the guiding center $\mathbf x$, the electric field $\mathbf E(\mathbf x^{\prime})$ has no component perpendicular to $\mathbf B(\mathbf x)$ and is determined as
\begin{equation}\label{Ez}
\begin{split}
E_\rho=0, \ \ \ E_\phi=0 \ \ \ \text{and} \ \ \ E_z=-\frac{\partial \varphi}{\partial z}.
\end{split}
\end{equation}
Assuming that $\mathbf B(\mathbf x^{\prime})$ is axially symmetric in the co-moving wind frame (i.e., $B_\phi=0$), and evaluating $\partial B_z/\partial z$ at the guiding center,  Maxwell's equation $\nabla_{\mathbf q} \cdot \mathbf B=0$ leads to
\begin{equation}\label{Brho}
B_\rho \approx-\frac{\rho }{2}\left.\frac{\partial  B_z}{\partial z}\right|_{\rho=0}.
\end{equation}
Upon substituting Eq.~(\ref{Brho}) into $\mathbf B=\nabla_{\mathbf q} \times \mathbf A$ and using the Coulomb gauge, we find 
\begin{equation}\label{vecpo}
A_\rho=0, \ \ \ A_\phi=\frac{\rho}{2}B_z(\mathbf x) \ \ \ \text{and} \ \ \ A_z=0.
\end{equation}
Following Eqs.~(\ref{lag}) through (\ref{genemoz}) and Eq.~(\ref{vecpo}), the Hamiltonian function is given from $\mathcal H=\mathbf p \cdot \dot{\mathbf q} - \mathcal L$ as
\begin{equation}\label{Ham}
\begin{split}
\mathcal H\!=\!\frac{p_\rho^2}{2m_e}\!+\!\frac{p_z^2}{2m_e}\!+\!\frac{1}{2m_e \rho^2} \!\left(p_\phi\!-\!\frac{q_e \rho A_\phi}{c}\right)^{\!\!2}\!+\!q_e\varphi.
\end{split}
\end{equation}
The motion of a particle is fully described by Hamilton's equations
\begin{equation}\label{cano}
\dot {\mathbf p}=-\frac{\partial \mathcal H}{\partial \mathbf q}
\end{equation}
and
\begin{equation}
    \dot {\mathbf q}=\frac{\partial \mathcal H}{\partial \mathbf p}.
\end{equation}
According to Eq.~(\ref{cano}), we identify the force terms 
\begin{equation}\label{dotprho}
\begin{split}
\dot p_\rho=\frac{1}{m_e\rho^3}\left [ p_\phi^2-\left ( \frac{q_e \rho^2}{2c}B_z \right )^{\!\!2}\right]
\end{split}
\end{equation}
and
\begin{equation}\label{dotpz}
\begin{split}
\dot p_z\!= \frac{q_e}{2m_ec}\!\left(\!p_\phi\!-\!\frac{q_e \rho^2}{2c}B_z\!\right)\hat {\mathbf b}\cdot \nabla_{\mathbf x} B_z+q_e E_z.
\end{split}
\end{equation}
where we use $B_{z}=B_{z}(\mathbf x)$, $E_{z}=E_{z}(\mathbf x)$, and
\begin{equation}\label{zetaderiv}
\frac{\partial B_z(\mathbf x)}{\partial z} \equiv \hat {\mathbf b}\cdot \nabla_{\mathbf x}B_z(\mathbf x).
\end{equation}
Because $\phi$ is a cyclic coordinate in Eq.~(\ref{Ham}), the azimuthal equation of motion is 
\begin{equation}\label{pphi}
\begin{split}
p_\phi=m_e \rho^2 \dot \phi  + \frac{q_e \rho^2}{2c}B_z=\text{const}.
\end{split}
\end{equation}
Without loss of generality, we choose our coordinate axis ($z$-axis) for a given particle so that $\rho$ is approximately constant  on the timescale of a few gyro-periods; i.e., the particle gyrates about the coordinate axis without quickly changing its $\rho$-coordinate.\footnote{It can be shown that such a coordinate system exists for each individual particle in a homogeneous magnetic field. Even if the magnetic field is inhomogeneous, this choice of coordinate system is still appropriate as long as the gyro-orbits are quasi-circular, which is fulfilled in the regime of adiabatic invariance.} With this choice,  $p_\phi$ from Eq.~(\ref{dotprho}) is
\begin{equation}\label{pphicon}
\begin{split}
p_\phi=-\frac{q_e \rho^2}{2c}B_z.
\end{split}
\end{equation}
Then, Eq.~(\ref{pphi}) leads to 
\begin{equation}\label{dotphi}
\begin{split}
\dot \phi=-\frac{q_eB_z}{m_e c},
\end{split}
\end{equation}
which is the electron cyclotron frequency. Using Eqs.~(\ref{dotprho}) through (\ref{dotphi}), the external forces in the $v_\parallel$- and $v_\perp$-directions are determined by 
\begin{equation}\label{dpal}
\dot v_\parallel=\frac{q_e E_z}{m_e}-\frac{v_\perp^2}{2 B_z}\hat {\mathbf b}\cdot \nabla_{\mathbf x} B_z
\end{equation}
and
\begin{equation}\label{dper}
\dot v_\perp=\frac{ v_\perp}{2 B_z}\mathbf v\cdot \nabla_{\mathbf x} B_z,
\end{equation}
where we use
\begin{equation}
\frac{d}{dt}=\mathbf v\cdot \nabla_{\mathbf x}
\end{equation}
as the total time derivative in the co-moving reference frame under steady-state conditions. Eqs.~(\ref{dpal}) and (\ref{dper}) describe the magnetic mirror force and the electric force exerted on all individual electrons in the co-moving solar-wind frame. 

\subsection{Kinetic Transport Equation and Moments}\label{2.4}

Substituting Eqs.~(\ref{dpal}) and (\ref{dper}) into Eq.~(\ref{gyrofulconvec}), we obtain
\begin{equation}\label{finalforce}
\begin{split}
&\frac{\partial f_e}{\partial t}\!+\!(\mathbf U\!+\!\langle \mathbf v \rangle_{\!\phi_v} )\!\cdot \!\nabla_{\mathbf x} f_e\!-\!\Big\langle\!\big[ (\mathbf v \! \cdot \!\nabla_{\mathbf x}) \mathbf U \big] \! \cdot \! \hat{\mathbf e}_\perp \!\Big\rangle_{\!\phi_v} \! \frac{\partial f_e}{\partial v_\perp}\\
&-\!\Big\{\big[(\mathbf U\!+\!\langle \mathbf v \rangle_{\!\phi_v} )\!\cdot\! \nabla_{\mathbf x}\big] \mathbf U \Big\}\! \cdot  \hat{\mathbf b} \frac{\partial f_e}{\partial v_\parallel}\\
&+\left(\frac{q_e E_z}{m_e}-\frac{v_\perp^2}{2 B_z}\hat {\mathbf b}\cdot \nabla_{\mathbf x} B_z \right)\frac{\partial f_e}{\partial v_\parallel}\\
&+\frac{ v_\perp}{2 B_z} \langle \mathbf v \rangle_{\!\phi_v}\cdot \nabla_{\mathbf x} B_z \frac{\partial f_e}{\partial v_\perp}=\left(\frac{\partial f_e }{\partial t}\right)_{\mathrm{col}}.
\end{split}
\end{equation}
Without loss of generality, we model the kinetic evolution of $f_e$ only at the origin of the co-moving wind frame:
\begin{equation}\label{fulconvecatorigin}
\left.\left\langle\frac{df_e}{dt}\right\rangle_{\!\!\phi_v}\right |_{z=0}=\left.\left(\frac{\partial f_e }{\partial t}\right)_{\mathrm{col}}\right |_{z=0}.
\end{equation}
We assume that $f_e$, $\mathbf U$, $\mathbf E$, and $\mathbf B$ are symmetric in $\Phi$ and $\Theta$. We note that $\langle\mathbf v \rangle_{\!\phi_v}=\hat{\mathbf b} v_\parallel$. We resolve the coefficient of the third term in Eq.~(\ref{finalforce}) by applying the coordinate choice that $\hat{\mathbf e}_x=\hat{\mathbf e}_R \sin\theta +\hat{\mathbf e}_\Phi \cos\theta$, $\hat{\mathbf e}_y=-\hat{\mathbf e}_\Theta$ (see Fig.~\ref{Fig1}a), and
\begin{equation}\label{Bgradi}
\begin{split}
    \frac{\partial \ln B_z}{\partial r}&=\frac{\partial \ln B_R}{\partial r}-\frac{\partial \ln \cos\theta}{\partial r}\\
    &=-\frac{2}{r}+\frac{\sin^2\theta}{r}-\frac{\sin^2\theta}{U}\frac{\partial U}{\partial r}.
\end{split}
\end{equation}
Then, Eq.~(\ref{fulconvecatorigin}) becomes
\begin{equation}\label{convecff}
\begin{split}
&\frac{\partial f_e}{\partial t}+\left(U+ v_{\parallel}\cos\theta  \right)\frac{\partial f_e}{\partial r}+\frac{q_e E_z}{m_e}\frac{\partial f_e}{\partial v_\parallel}\\
&-\left[\left(U+v_{\parallel}\cos\theta  \right)\cos\theta\frac{\partial U }{\partial r} +v_\parallel \sin^2\theta \frac{U}{r} \right ]\frac{\partial f_e}{\partial v_\parallel}\\
&+ \frac{v_\perp}{2}\frac{\partial \ln B_z}{\partial r}\!\left [\!\left(U+v_\parallel\cos\theta \right )\!\frac{\partial f_e}{\partial v_\perp}- v_\perp \cos\theta \frac{\partial f_e}{\partial v_\parallel}  \right]\\
&=\left(\frac{\partial f_e}{\partial t}\right)_{\mathrm{col}},
\end{split}
\end{equation}
where $f_e\equiv f_e(r,v_\perp,v_\parallel,t)$, $U \equiv U(r)$, $\theta \equiv \theta(r)$, $E_z\equiv E_z(r)$ and $B_z \equiv B_z(r)$. Eq.~(\ref{convecff}) is our kinetic transport equation for the kinetic evolution in the co-moving frame shown  in Fig.~\ref{Fig1}b. Considering Eq.~(\ref{Bgradi}), our Eq.~(\ref{convecff}) is equivalent to equation (15) by \citet{2007ApJ...662..350L} after transforming Eq.~(\ref{convecff})  from cylindrical to spherical velocity coordinates. Likewise, our  Eq.~(\ref{convecff}) is consistent with the transport equations derived by \citet{1966AnPhy..37..487L}, \citet{1971ApJ...170..265S}, \citet{1985ApJ...296..319W}, \citet{https://doi.org/10.1029/96JA03671}, \citet{le_Roux_2009}, and \citet{book}.

Assuming that Coulomb collisions do not change the number of particles, we confirm that the zeroth moment of Eq.~(\ref{convecff}) is equivalent to the continuity equation in spherical coordinates with $\Phi$- and $\Theta$-symmetry:
\begin{equation}\label{continuity}
\begin{split}
\frac{\partial n_e}{\partial t}+\frac{1}{r^2}\frac{\partial [r^2 n_e(U+\overline{U})]}{\partial r}=0,
\end{split}
\end{equation}
where
\begin{equation}\label{updatene}
\begin{split}
n_e=\int f_e d^3\mathbf v,
\end{split}
\end{equation}
and
\begin{equation}\label{bulkspeed}
\begin{split}
\overline{U}=\frac{1}{n_e}\int v_\parallel f_e d^3\mathbf v.
\end{split}
\end{equation}
The $\overline{U}$-term in Eq.~(\ref{continuity}) arises only if our system develops a non-zero bulk velocity in the co-moving reference frame. Our numerical approach guarantees that $\overline U=0$ when the system is in steady state. In such a steady-state configuration leading to $\overline{U}=0$, the particle flux $r^2 n_eU$ is conserved. 

We determine the self-consistent electric field by taking the first moment of Eq.~(\ref{convecff}) and re-arranging the resulting expression to
\begin{equation}\label{efield}
\begin{split}
&E_z(r)=\frac{m_e}{q_en_e}\frac{\partial (n_e \overline{U})}{\partial t}+\cos\theta\frac{k_B}{q_e n_e }\frac{\partial (n_e T_{\parallel e})}{\partial r}\\
&-\cos\theta\frac{k_B}{q_e}\frac{\partial \ln B_z}{\partial r}(T_{\parallel e}-T_{\perp e})\\
&+\frac{m_e U}{q_e}\cos\theta\frac{\partial U}{\partial r}-\frac{m_e}{q_en_e}\int v_\parallel \left( \frac{\partial f_e}{\partial t} \right)_{\mathrm{col}} d^3\mathbf v,
\end{split}
\end{equation}
where 
\begin{equation}\label{updateTepal}
\begin{split}
T_{\parallel e}(r)=\frac{m_e}{k_B n_e}\int v_\parallel^2 f_e d^3\mathbf v,
\end{split}
\end{equation}
\begin{equation}\label{updateTeper}
\begin{split}
T_{\perp e}(r)=\frac{m_e}{2k_B n_e}\int v_\perp^2 f_e d^3\mathbf v,
\end{split}
\end{equation}
and $k_B$ is the Boltzmann constant.
Eq.~(\ref{efield}) is the same as the generalized Ohm's law based on the electron fluid equation of motion under our assumptions \citep{1997JGR...102.4701L,refId0}. As our system relaxes to a steady state, the first term in Eq.~(\ref{efield}) disappears (i.e., Eq.~(\ref{bulkspeed}) is zero). The last term of Eq.~(\ref{efield}) corresponds to the thermal force by Coulomb collisions \citep{2019ApJ...882..146S}. We evaluate the integral in this term numerically.

\subsection{Coulomb Collisions}\label{2.5}

In order to model the scattering through Coulomb collisions, we adopt the Fokker--Planck operator given by \citet{doi:10.1098/rspa.1990.0026} and \citet{Vocks_2002} with Rosenbluth potentials \citep{PhysRev.107.1}:
\begin{equation}\label{FF}
\begin{split}
&\left(\frac{\partial  f_e }{\partial  t}\right)_{\mathrm{col}}
=\sum_{b} \Gamma_{eb} \left (4\pi \frac{m_e}{m_b} f_b  f_e +\frac{\partial H_b}{\partial v_\perp} \frac{\partial f_e}{\partial v_\perp}\right. \\
&+\frac{\partial H_b}{\partial v_\parallel} \frac{\partial f_e}{\partial v_\parallel}+ \frac{1}{2} \frac{\partial^2 G_b}{\partial v_\perp^2}  \frac{\partial^2 f_e}{\partial v_\perp^2}+\frac{1}{2} \frac{\partial^2 G_b}{\partial v_\parallel^2}  \frac{\partial^2 f_e}{\partial v_\parallel^2}\\
&\left.+\frac{\partial^2 G_b}{\partial v_\perp \partial v_\parallel}  \frac{\partial^2 f_e}{\partial v_\perp \partial v_\parallel}+\frac{1}{2v_\perp^2}\frac{\partial G_b}{\partial v_\perp}\frac{\partial f_e}{\partial v_\perp}\right ), 
\end{split}
\end{equation}
where
\begin{equation}\label{RosenG}
G_b(\mathbf x,\mathbf v)\equiv\int f_b (\mathbf x,\mathbf v^\prime)|\mathbf v-\mathbf v^\prime|d^3 \mathbf v^\prime,
\end{equation}
\begin{equation}\label{RosenH}
H_b(\mathbf x,\mathbf v)\equiv\frac{m_b-m_e}{m_b} \int f_b (\mathbf x,\mathbf v^\prime)|\mathbf v-\mathbf v^\prime|^{-1}d^3 \mathbf v^\prime,
\end{equation}
and
\begin{equation}\label{cp}
\Gamma_{eb}\equiv4\pi\bigg ( \frac{Z_b q_e^2}{m_e} \bigg )^2  \ln \Lambda_{eb}.
\end{equation}
The subscript $b$ indicates the background particle species. The quantity $\ln\Lambda_{eb}$ is the Coulomb logarithm. We set it to a constant value of $\ln \Lambda_{eb}\approx 25$, which is typical for space plasmas. The parameter $Z_b$ is the charge number of a particle of species $b$. For the background VDFs, we only consider electrons and protons and assume that the background electron and proton VDFs are gyrotropic and Maxwellian:
\begin{equation}\label{max}
    f_b(r,v)=\frac{n_b }{\pi^{3/2} v_{th,b}^3}\exp\left( -\frac{v^2}{v_{th,b}^2} \right),
\end{equation}
where $v_{th,b}(r) \equiv \sqrt{2k_{B}T_{b}(r)/m_b}$,  $v^2=v_\perp^2+v_\parallel^2$, $n_b(r)$ is the density and $T_b(r)$ is the temperature of the background particles at distance $r$. Then, the Rosenbluth potentials Eqs.~(\ref{RosenG}) and (\ref{RosenH}) yield
\begin{equation}\label{G}
G_b(r,v)= \pi v_{th,b}^4 f_b+n_b\frac{v_{th,b}^2+2v^2}{2v}\erf \left(\frac{v}{v_{th,b}} \right),
\end{equation}
and
\begin{equation}\label{H}
H_b(r,v)=\frac{m_b-m_j}{m_b}\frac{n_b}{v}\erf \left(\frac{v}{v_{th,b}} \right),
\end{equation}
where $\erf(x)$ is the error function. For numerical reasons, we apply a Taylor expansion for $(v_{th,b}/v)\erf (v/v_{th,b})$ at $v/v_{th,b}=0$ to Eqs.~(\ref{G}) and (\ref{H}) if $v/v_{th,b}<1$.

In our study, we assume that the temperatures of the background electrons and the background protons are equal in our region of interest. The proton-to-electron temperature ratio varies in the solar wind and is generally close to unity only when collisions are sufficient to equilibrate the temperatures \citep{Cranmer_2020,d437a059c4a9467cbb87b86c0f01b676}. However, this temperature ratio has only a small impact on our results since it enters our calculation solely via the collision operator.

\section{Numerical Treatment of the Kinetic Transport Equation} \label{3}

As a first step, we aim to evaluate our model near the Sun at distances (from $r/r_s=5$ to $r/r_s=20$), at which direct in-situ measurements of the electron VDF are missing. Even though we only evaluate our kinetic equation Eq.~(\ref{convecff}) near the Sun in the present paper, it is generally valid also at greater distances from the Sun where the effect of the spiral field geometry is greater.

In Appendix~\ref{A}, we present our mathematical approach to numerically solve Eq.~(\ref{convecff}) based on a combination of a Crank--Nicolson scheme in velocity space and an Euler scheme in radial space. Our numerical solution given in Eq.~(\ref{sol}) implements the time evolution of the electron VDF as a function of $r$, $v_\perp$ and $v_\parallel$.

\subsection{Overall Numerical Strategy} \label{3.1}
We normalize $r$ by  $r_s$, and $v_\perp$ and $v_\parallel$ by  the electron Alfv\'en velocity estimated at 1\,au, denoted as $v_{Ae0}$. We  consequently normalize time $t$ in units of $r_s/v_{Ae0}$. We define the discrete electron VDF as $f_{L,M,N}^{T}\equiv f_e(r_L, \mathit{v}_{\perp M},\mathit{v}_{\parallel N}, t^T )$, where the radial index $L$ counts as 1, 2, \dots, $N_r$, the velocity indexes $M$ and $N$ count as 1, 2, \dots, $N_v$, and the time index $T$ counts as 1, 2, \dots (see also Appendix~\ref{A}). We iterate the calculation of our numerical solution according to Eq.~(\ref{sol}) until the $2$-norm of the residual difference in subsequent VDFs,
\begin{equation}\label{norm}
\begin{split}
\big\| f^{T+1}_{L,M,N}&- f^{T}_{L,M,N}\big\|_2\\
&\!\!\!\!\!\!\!\!=\left[\sum_{L=1 }^{N_r} \sum_{M=1 }^{N_v} \sum_{N=1 }^{N_v} \big|  f^{T+1}_{L,M,N}- f^{T}_{L,M,N} \big|^2 \right]^{1/2},
\end{split}
\end{equation}
reaches a minimum value, which we identify with a quasi-steady state. After reaching this quasi-steady state, we only analyze the dependence of $f^{T}_{L,M,N}$ on $r$, $v_\perp$, and $v_\parallel$.

During our calculation, we update the  electric field via Eq.~(\ref{efield}) and the collisional background species parameters in Eq.~(\ref{max}) every 40 time steps through Eqs.~(\ref{updatene}), (\ref{updateTepal}), (\ref{updateTeper}) and
\begin{equation}\label{totaltemp}
T_{e}=\frac{2T_{\perp e}+T_{\parallel e}}{3}.
\end{equation}

\subsection{Initial Conditions} \label{3.2}

At the coronal lower boundary of our simulation domain, collisions are sufficient to create a Maxwellian thermal core of the electrons. However, non-thermal tails can already exist at $r=5r_s$. Non-Maxwellian electron distributions in the corona are often evoked in kinetic models of the solar wind \citep{1992ApJ...398..319S,1992ApJ...398..299S,1997A&A...324..725M,2000ApJ...528..509V}. Therefore,  we select a $\kappa$-distribution with $\kappa=8$ for the initial electron VDF  \citep{https://doi.org/10.1029/2009JA014352,2013SSRv..175..183L,2016Ap&SS.361..359N}:
\begin{equation}\label{input}
\begin{split}
f_e=\frac{n_e}{ v_{th,e}^3 }\left[\frac{2}{\pi (2\kappa-3)}\right]^{3/2}&\frac{\Gamma(\kappa+1)}{\Gamma(\kappa-0.5)}\\
&\!\!\!\!\!\!\!\!\!\!\!\!\!\!\!\!\times \left (1+\frac{2}{2\kappa-3}\frac{v^2}{ v_{th,e}^2} \right)^{-\kappa-1},
\end{split}
\end{equation}
where $\Gamma(x)$ is the $\Gamma$-function and $\kappa>3/2$ is the $\kappa$-index. 

Even though there is a wide range of natural variation, we prescribe representative initial profiles in our region of interest for the  bulk speed as \citep{Bemporad_2017,YAKOVLEV2018552}
\begin{equation}\label{windspeed}
U(r)=(400\,\mathrm{km/s}) \tanh\left({\frac{r}{10r_s}}\right),
\end{equation}
for the  electron density as
\begin{equation}\label{density}
n_e(r)=(5\,\mathrm{cm}^{-3})  \left (\frac{215r_s}{r} \right)^2\left (\frac{U(r=215r_s)}{U(r)} \right),
\end{equation}
and for the electron temperature as \citep{1989JGR....94.6893M,Moncuquet_2020}
\begin{equation}\label{temperature}
T_e(r)=(10^6\,\mathrm K) \left (\frac{5r_s}{r} \right)^{0.8}.
\end{equation}
In Eq.~(\ref{density}), the factor $U(r=215r_s)/U(r)$ guarantees  mass-flux conservation under steady-state conditions (i.e., $r^2 n_e U=\mathrm{const}$). We assume that the profile of $U(r)$ stays constant during our calculation. Thus, $U(r)$ starts from 185\,km/s at the inner boundary and reaches 385\,km/s at the outer boundary of our integration domain. However, the profiles of $n_e$ and $T_e$ evolve through the evolution of $f_e$. By applying Eqs.~(\ref{windspeed}) through (\ref{temperature}) to Eq.~(\ref{input}), we initially define our electron VDF at all radial distances. We also initially apply Eqs.~(\ref{windspeed}) through (\ref{temperature}) to Eq.~(\ref{max}) for the collisional background species. For the background magnetic field in Eq.~(\ref{Bz}), we set
$B_0=0.037\,\mathrm{G}$ and $r_0=5r_s$ based on the PSP measurements presented by \cite{2021A&A...650A..18B}. We focus our analysis on the equatorial heliospheric plane (i.e., $\Theta=90^\circ$). We then calculate $v_{Ae0}=836\,\mathrm{km/s}$ by using Eq.~(\ref{Bz}) and Eq.~(\ref{density}). In our calculation, the radial step size is $\Delta r/r_s=0.25$, the step size in velocity space is $\Delta v/v_{Ae0}=0.45$, and the size of the time step is $\Delta t/(v_{Ae0}/r_s)=1.2\times10^{-3}$.

\subsection{Boundary Conditions and Smoothing in Velocity Space} \label{3.3}
For the boundary conditions in velocity space, we first estimate the ratios between adjacent VDFs in $v_\parallel$ and $v_\perp$  as
\begin{equation}\label{velocityratio1}
\Upsilon_{\parallel L,M,N}^{T}=\frac{f_{L,M,N}^{T}}{f_{L,M,N+1}^{T}}
\end{equation} 
and
\begin{equation}\label{velocityratio2}
\Upsilon_{\perp L,M,N}^{T}=\frac{f_{L,M,N}^{T}}{f_{L,M+1,N}^{T}}.
\end{equation} 
We then update the VDF values at the given boundary in each time step by using Eqs.~(\ref{velocityratio1}) and (\ref{velocityratio2}), evaluated at the previous time step, as
\begin{equation}\label{velocityboundary}
f_{L,M,1}^{T}=f_{L,M,2}^{T}\Upsilon_{\parallel L,M,1}^{T-1},
\end{equation}
\begin{equation}
f_{L,M,N_v}^{T}=\frac{f_{L,M,N_v-1}^{T}}{\Upsilon_{\parallel L,M,N_v-1}^{T-1}},
\end{equation}
\begin{equation}
f_{L,1,N}^{T}=f_{L,2,N}^{T}\Upsilon_{\perp L,1,N}^{T-1},
\end{equation}
and
\begin{equation}
f_{L,N_v,N}^{T}=\frac{f_{L,N_v-1,N}^{T}}{\Upsilon_{\perp L,N_v-1,N}^{T-1}}.
\end{equation}

To avoid numerical errors caused by our limited velocity resolution, we apply an averaging scheme to smooth the VDFs in velocity space. We average $f_{L,M,N}^{T}$ in each time step by using Eqs.~(\ref{velocityratio1}) and (\ref{velocityratio2}), evaluated at the previous time step, as
\begin{equation}\label{velocitysmooth}
\begin{split}
&\left<f\right>_{L,M,N}^{T}\!= \!\frac{f_{L,M,N+1}^{T}\Upsilon_{\parallel L,M,N+1}^{T-1}}{8}\!+\!\frac{f_{L,M,N-1}^{T}}{8\Upsilon_{\parallel L,M,N-1}^{T-1}}\\
&+\!\frac{f_{L,M,N}^{T}}{2}\!+\!\frac{f_{L,M+1,N}^{T}\Upsilon_{\perp L,M,N}^{T-1}}{8}\!+\!\frac{f_{L,M-1,N}^{T}}{8\Upsilon_{\perp L,M-1,N}^{T-1}},
\end{split}
\end{equation}
where we denote the averaged VDF as $\left<f\right>_{L,M,N}^{T}$. The approach described by Eq.~(\ref{velocitysmooth}) improves the numerical stability without changing the  physics of the model. In Appendix~\ref{b}, we show the result of our model without smoothing for comparison.

\subsection{Boundary Conditions and Smoothing in Configuration Space} \label{3.4}
For the outer boundary in our $r$-coordinate, we first estimate the ratio between radially adjacent VDFs at the initial time step $T=1$ as
\begin{equation}\label{ratio}
\begin{split}
\Upsilon_{L,M,N}=\frac{f_{L,M,N}^{1}}{f_{L+1,M,N}^{1}}.
\end{split}
\end{equation} 
We then update the VDF at the outer boundary in each time step by using Eq.~(\ref{ratio}) as \begin{equation}\label{radialboundarycondition}
\begin{split}
f_{N_r,M,N}^{T}=\frac{f_{N_r-1,M,N}^{T}}{\Upsilon_{N_r-1,M,N}}.
\end{split}
\end{equation}
This corresponds to an open outer boundary condition at $L=N_r$.

The corona is so collisional that the assumption of a constant isotropic VDF at the inner boundary at $r=5r_s$ is reasonable. Thus, we require that our initial VDF at the inner boundary, denoted as $f_{1,M,N}^{T}$, remains constant throughout our calculation. This choice of a constant inner boundary in conjunction with the large radial gradients at small $r$ can lead to a fast growth of numerical errors. In order to compensate for these errors, we apply an averaging scheme to smooth the VDF in configuration space, following a similar scheme as described in Section~\ref{3.3} for velocity space. We average $f_{L,M,N}^{T}$ (except for $f_{1,M,N}^{T}$ and $f_{N_r,M,N}^{T}$) in each time step by using Eq.~(\ref{ratio}) as
\begin{equation}\label{average}
\begin{split}
\left<f\right>_{L,M,N}^{T}= \frac{f_{L-1,M,N}^{T}}{4\Upsilon_{L-1,M,N}}&+\frac{f_{L,M,N}^{T}}{2}\\
&+\frac{f_{L+1,M,N}^{T}\Upsilon_{L,M,N}}{4}.
\end{split}
\end{equation}
We apply the averaging in configuration space before the averaging in velocity space.

\section{Results} \label{4}

\subsection{Kinetic Expansion} \label{4.1}

\begin{figure}[ht]
  \centering
  \includegraphics[width=\columnwidth]{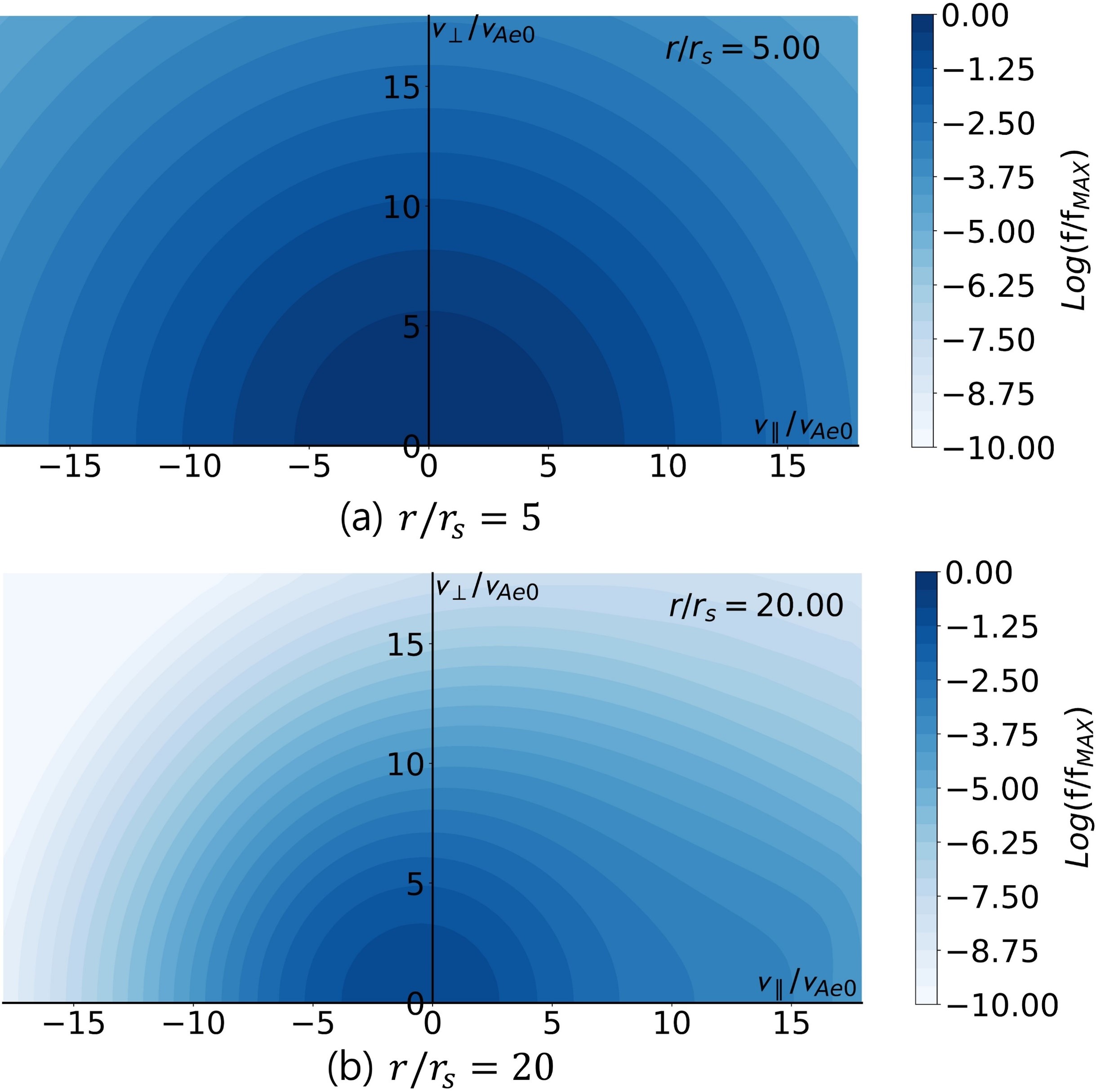}
  \caption{Kinetic evolution of the electron VDF from (a) $r/r_s=5$ to (b) $r/r_s=20$. The value of the distribution function is normalized to the maximum value of the VDF at the inner boundary. The electron density decreases due to  spherical expansion. The effects of  particle streaming and the magnetic mirror force mainly contribute to the formation of the electron strahl at  positive $v_\parallel$ and small $v_{\perp}$. Because of the electric field, the electron core slightly shifts towards negative $v_\parallel$. An animation of this figure is available. The animation shows the kinetic evolution of the electron VDF from $r/r_s=5$ to $r/r_s=20$. Panels (a) and (b) show the initial and final snapshots of the animation. }
  \label{2Dresults}
\end{figure}

\begin{figure}[ht]
  \centering
  \includegraphics[width=0.45\textwidth]{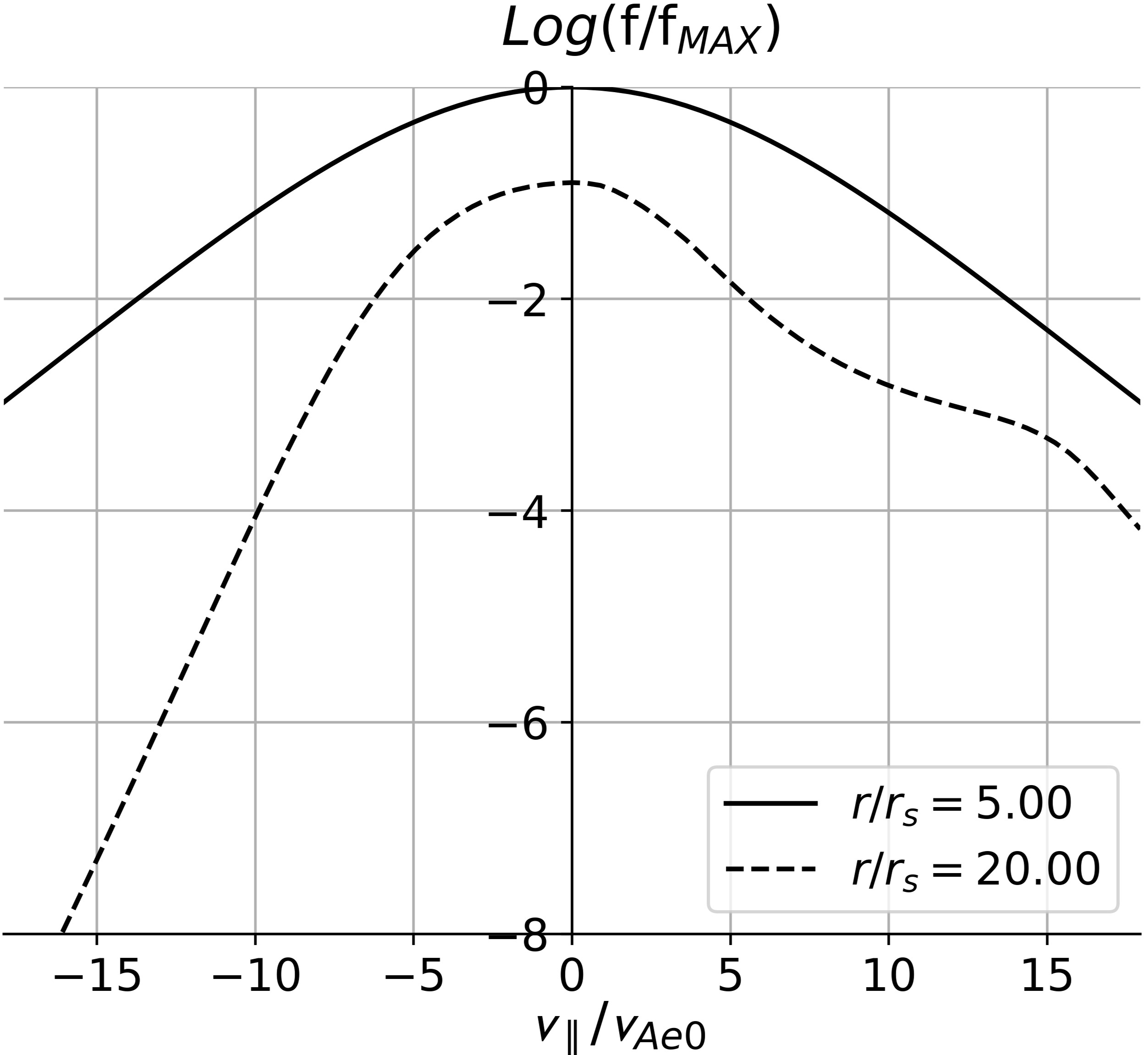}
  \caption{ Kinetic evolution of the electron VDF as cuts in the $v_\parallel$-direction. An animation of this figure is available. The animation shows the evolution of the cuts in the $v_{\parallel}$-direction  from $r/r_s=5$ to $r/r_s=20$. The figure shows the initial and final snapshots of the animation.}  
  \label{1Dresults}
\end{figure}

Following the numerical treatment discussed in Section~\ref{3}, we acquire the results for the kinetic evolution of the electron VDF from $r/r_s=5$ to $r/r_s=20$ according to Eq.~(\ref{convecff}). We show the two-dimensional electron VDF both at $r/r_s=5$ and at $r/r_s=20$ in Fig.~\ref{2Dresults}.
Fig.~\ref{1Dresults} shows one-dimensional cuts along the $v_\parallel$-direction of the same distributions shown in Fig.~\ref{2Dresults}. Fig.~\ref{2Dresults}a and the black solid curve in Fig.~\ref{1Dresults} show the electron VDF at $r/r_s=5$.  Fig.~\ref{2Dresults}b and the black dashed curve in Fig.~\ref{1Dresults} show the electron VDF at $r/r_s=20$. Animations of these figures are available in the supplementary material. This kinetic evolution is the result from the combined effects of the accelerating solar wind, particle streaming, the magnetic mirror force, the electric field, the geometry of the Parker-spiral magnetic field, and Coulomb collisions throughout the spherical expansion.

\begin{figure}[ht]
  \centering
  \includegraphics[width=0.45\textwidth]{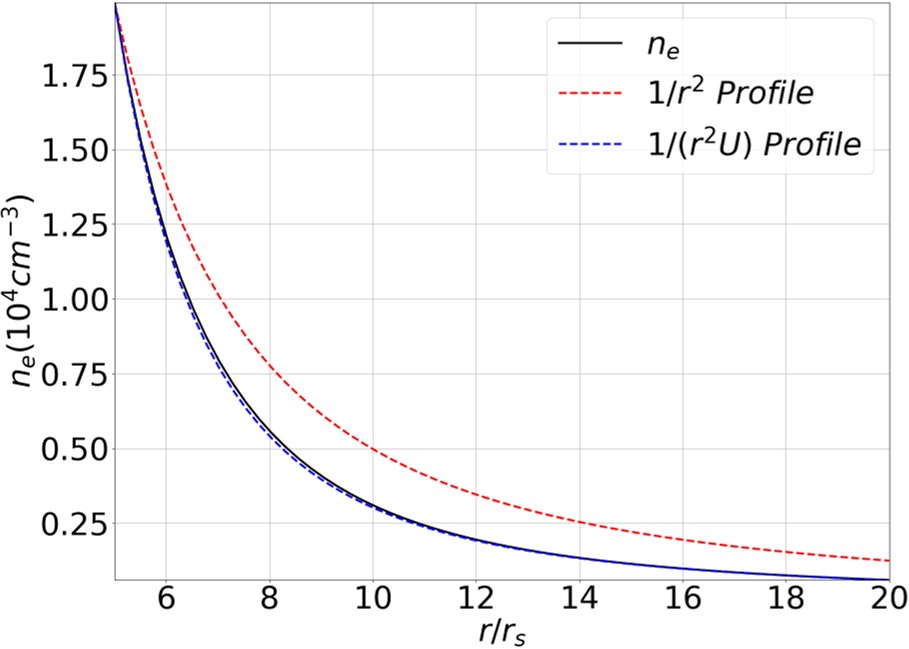}
  \caption{Profiles of the electron density as a function of radial distance. The black solid curve shows the density as calculated from the electron VDFs with Eq.~(\ref{updatene}). The blue dashed curve is the initial profile of the electron density given by Eq.~(\ref{density}).  The red dashed curve shows a $1/r^2$-profile for reference.}
  \label{actualdensity}
\end{figure}

In Fig.~\ref{actualdensity}, the black solid curve represents the radial density profile calculated with Eq.~(\ref{updatene}). Both Figs.~\ref{2Dresults} and \ref{1Dresults} illustrate that the electron density decreases from $1.9\times 10^{4}\,\mathrm{cm}^{-3}$ to $0.1\times 10^{4}\,\mathrm{cm}^{-3}$ in our region of interest as a consequence of the spherical expansion. The blue dashed curve in Fig.~\ref{actualdensity} is our initial density profile according to Eq.~(\ref{density}). The difference between both profiles shows that, during our calculation, the density profile of the electron VDF stays nearly constant to satisfy Eq.~(\ref{continuity}) under  steady-state conditions as expected. The red dashed curve represents a $1/r^2$-profile for comparison. Because the solar wind still undergoes acceleration in our model domain according to Eq.~(\ref{windspeed}), the electron density decreases faster with distance than the $1/r^2$-profile (see the terms in the second line of Eq.~(\ref{convecff})).

Electrons with $(U+v_{\parallel}\cos\theta)<0$ stream into the sunward direction, while electrons with $(U+v_\parallel\cos\theta )>0$ stream into the anti-sunward direction. Electrons with $(U+v_\parallel\cos\theta )>0$ are continuously supplied from the Sun to our system. The streaming, in combination with our radial temperature gradient given in Eq.~(\ref{temperature}), causes a deformation of the VDF with time according to the second term in the first line of Eq.~(\ref{convecff}). Therefore, the streaming effect contributes to the creation of the electron strahl above around $v_\parallel/v_{Ae0}=10$. Because of this deformation of the VDF, the magnetic mirror force (the terms in the third line of Eq.~(\ref{convecff})) becomes more effective in the anti-sunward direction and focuses the electrons towards smaller $v_\perp$ at $(U+v_\parallel\cos\theta )>0$. 

The  electric field contributes with a sunward acceleration to the electron bulk motion. This effect moves  the center of the electron core, whose thermal energy is well below the electric potential energy, toward negative $v_\parallel$ as the solar wind expands (see the third term in the first line of Eq.~(\ref{convecff})). Moreover, the electric field guarantees that the bulk speed in the co-moving wind frame, Eq.~(\ref{bulkspeed}), stays at a value of zero. This situation means that our system always fulfils $\overline U=0$. At small velocities in Fig.~\ref{2Dresults}, the electrons are isotropic, which is the core part of the VDF. This core isotropy is due to the Coulomb collisions that isotropize more efficiently at small velocities.

\subsection{Electron VDF Fits}\label{4.2}

We apply fits to our numerical results for the electron VDF to quantify the core and strahl parameters and to compare our results with observations. Our fit routine uses the Nelder--Mead method in logarithmic space. We fit our electron VDFs with the sum of two bi-Maxwellian distributions for the electron core and strahl \citep{doi:10.1029/2008JA013883}:
\begin{equation}\label{fitVDF}
f_{\mathrm{fit}}=f_c+f_s,
\end{equation}
where
\begin{equation}\label{fitVDFc}
\begin{split}
f_c\!=\!\frac{n_c }{\pi^{3/2} v_{\perp th,c}^2v_{\parallel th,c}}\exp\!\left[ -\frac{v_\perp^2}{v_{\perp th,c}^2}\!-\!\frac{(v_\parallel\!-\!U_c)^2}{v_{\parallel th,c}^2} \right],
\end{split}
\end{equation}
\begin{equation}\label{fitVDFs}
\begin{split}
f_s\!=\!\frac{n_s }{\pi^{3/2} v_{\perp th,s}^2v_{\parallel th,s}}\exp\!\left[ -\frac{v_\perp^2}{v_{\perp th,s}^2}\!-\!\frac{(v_\parallel\!-\!U_s)^2}{v_{\parallel th,s}^2} \right],
\end{split}
\end{equation}
$v_{\perp th,j}\equiv \sqrt{2k_{B}T_{\perp j}/m_e}$, $v_{\parallel th,j}\equiv \sqrt{2k_{B}T_{\parallel j}/m_e}$, and the subscript $j$ indicates each electron population ($j=c$ for the core and $j=s$ for the strahl). In our fit parameters, we set $n_s=n_e-n_c$, where $n_c$ and $n_s$ are the core and strahl densities, $T_{\perp c}$ ($T_{\parallel c}$) and $T_{\perp s}$ ($T_{\parallel s}$) are the perpendicular (parallel) temperatures of core and strahl, and $U_c$ and $U_s$ are the bulk velocities of the core and the strahl.
\begin{figure}[ht]
  \centering
  \includegraphics[width=\columnwidth]{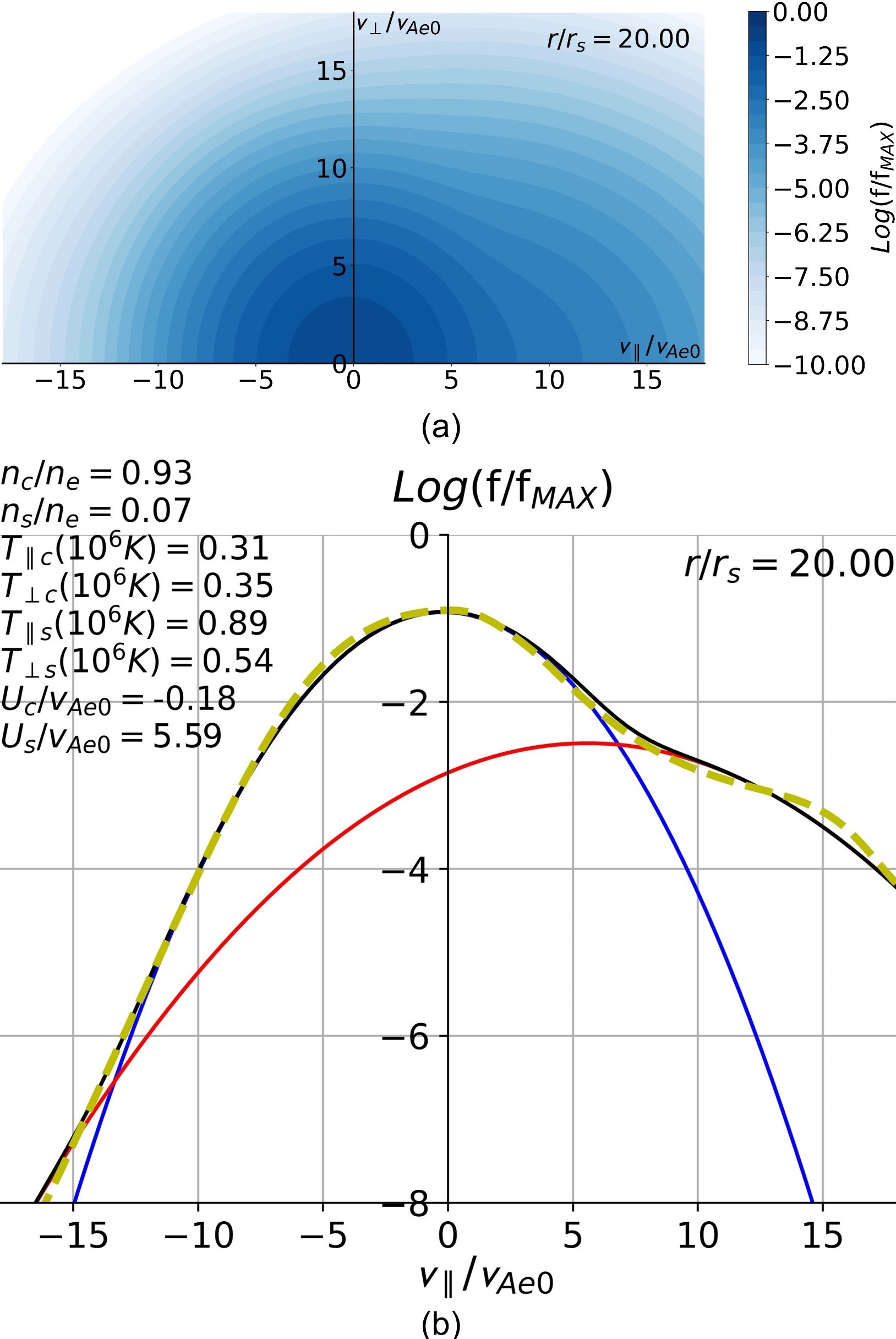}
  \caption{Fit results for the electron VDF. (a) Two-dimensional fit result at $r/r_s=20$. The input VDF is the simulation result shown in Fig.~\ref{2Dresults}b; (b) one-dimensional cut of the electron VDF in the $v_\parallel$-direction. The fit parameters are given in the top-left corner. The blue and red solid curves are the fit results for the core and strahl, respectively. The black solid curve is the sum of both fits according to Eq.~(\ref{fitVDF}), and the yellow dashed curve is the same as the dashed curve in Fig.~\ref{1Dresults}. Animations of these figures are available. The animations show the kinetic evolution of the fitted electron VDF in 2D ($v_\parallel$, $v_\perp$)-space and 1D $v_\parallel$-space from $r/r_s=5$ to $r/r_s=20$ in the same format as Figs.~\ref{2Dresults} and~\ref{1Dresults}.}
  \label{FittedVDF}
\end{figure}

Fig.~\ref{FittedVDF}a shows our fit result for the electron VDF  at $r/r_s=20$, for which Fig.~\ref{2Dresults}b shows the underlying direct numerical output. Fig.~\ref{FittedVDF}b shows the corresponding one-dimensional cut of the VDF in the  $v_\parallel$-direction.  In the top-left corner of  Fig.~\ref{FittedVDF}b, we provide the fit parameters from our analysis. The blue and red solid curves show the fitted VDFs for the core and strahl, respectively. The black solid curve shows the total $f_{\mathrm{fit}}$, and the yellow dashed curve is our direct numerical result; i.e., the same as the black dashed curve in Fig.~\ref{1Dresults}. We fit the electron VDFs at all radial distances from our numerical results, and an animation for the kinetic evolution of the fitted VDFs as a function of distance is available in the supplementary material. After fitting all electron VDFs, the normalized sum of squared residuals  is always less than 0.013, which quantifies the good agreement between our numerical results and the fit results \citep{Abraham2021_submitted}.
\begin{figure}[ht]
  \centering
  \includegraphics[width=\columnwidth]{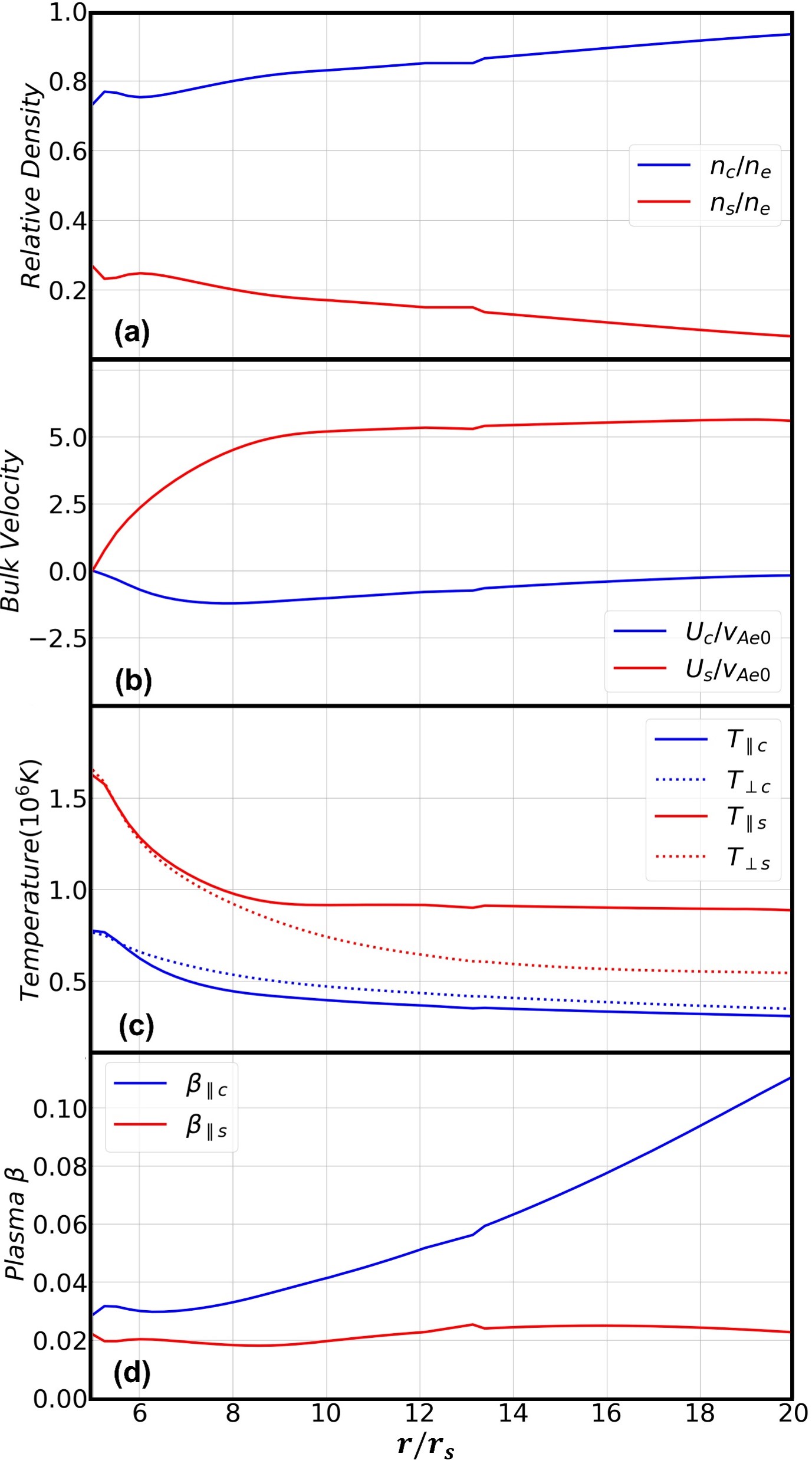}
  \caption{Radial profiles from our fit results of (a) the relative densities, where $n_e=n_c+n_s$; (b) the population bulk velocities; (c) their temperatures; and (d) $\beta_{\parallel c}$ and $\beta_{\parallel s}$. The blue curves show core parameters, and the red curves show  strahl parameters.}
  \label{fittedplots}
\end{figure}

Fig.~\ref{fittedplots} shows radial profiles for a selection of fit parameters from our model. The blue and red curves in each profile correspond to the core and strahl fit parameters, respectively. According to Fig.~\ref{fittedplots}a, the relative density of the electron strahl near the inner boundary is around 26\% of the total electron density. Such a high percentage is mostly a numerical artifact from our fitting scheme, which occurs whenever  the fitted core and strahl VDFs largely overlap because of a small $U_c$ and $U_s$ \citep{https://doi.org/10.1029/2005JA011119,doi:10.1029/2008JA013883}. At larger distances, the relative strahl density decreases continuously towards 7\% at the outer boundary. 

According to Fig.~\ref{fittedplots}b, $U_s$ increases from the inner boundary and then stays nearly constant at around $U_s/v_{Ae0}=5.6$ for $r/r_s\gtrsim 15$. At the same time, $U_c$ decreases near the inner boundary as a consequence of the strong  electric field near the corona.

As shown in Fig.~\ref{fittedplots}c, $T_{\parallel s}$ decreases rapidly near the inner boundary, and then stays nearly constant at around $9\times10^5\,\mathrm K$ at larger distances. On the other hand, $T_{\perp s}$, $T_{\perp c}$ and $T_{\parallel c}$ steadily decrease. We find that  $T_{\perp c}>T_{\parallel c}$ in all of our fit results (except for the inner boundary, where $T_{\perp c}=T_{\parallel c}$). 

Fig.~\ref{fittedplots}d shows the ratio between the parallel thermal  pressure to the magnetic-field pressure, $\beta_{\parallel j}=8\pi n_jk_BT_{\parallel j}/B_z^2$ separately for the core and for the strahl population. Both $\beta_{\parallel c}\ll 1$ and $\beta_{\parallel s}\ll 1$. We find that $\beta_{\parallel s}$ stays approximately constant at around 0.02 while $\beta_{\parallel c}$ steadily increases with distance.

\subsection{Comparison with PSP Data} \label{4.3}

We compare our numerical results with measurements from the dataset by \cite{Abraham2021_submitted}. This dataset is based on fits to the observed level-3 electron distributions provided by the Solar Wind Electron Alphas and Protons (SWEAP) instrument suite \citep{2016SSRv..204..131K,2020ApJS..246...74W} on board PSP. The analysis method fits bi-Maxwellian distributions to the core and strahl and a bi-$\kappa$-distribution to the halo. Because data below 30\,eV are contaminated with secondary electrons, the dataset neglects all measurement points below 30\,eV. In the \citet{Abraham2021_submitted} dataset, most VDFs ($\sim$4200 in total) can be fully modelled with bi-Maxwellian core and strahl VDFs, without the need to include a halo distribution in the range of radial distances between $r/r_s=20.3$ and $r/r_s=21.3$. 

\begin{figure}[ht]
  \centering
  \includegraphics[width=\columnwidth]{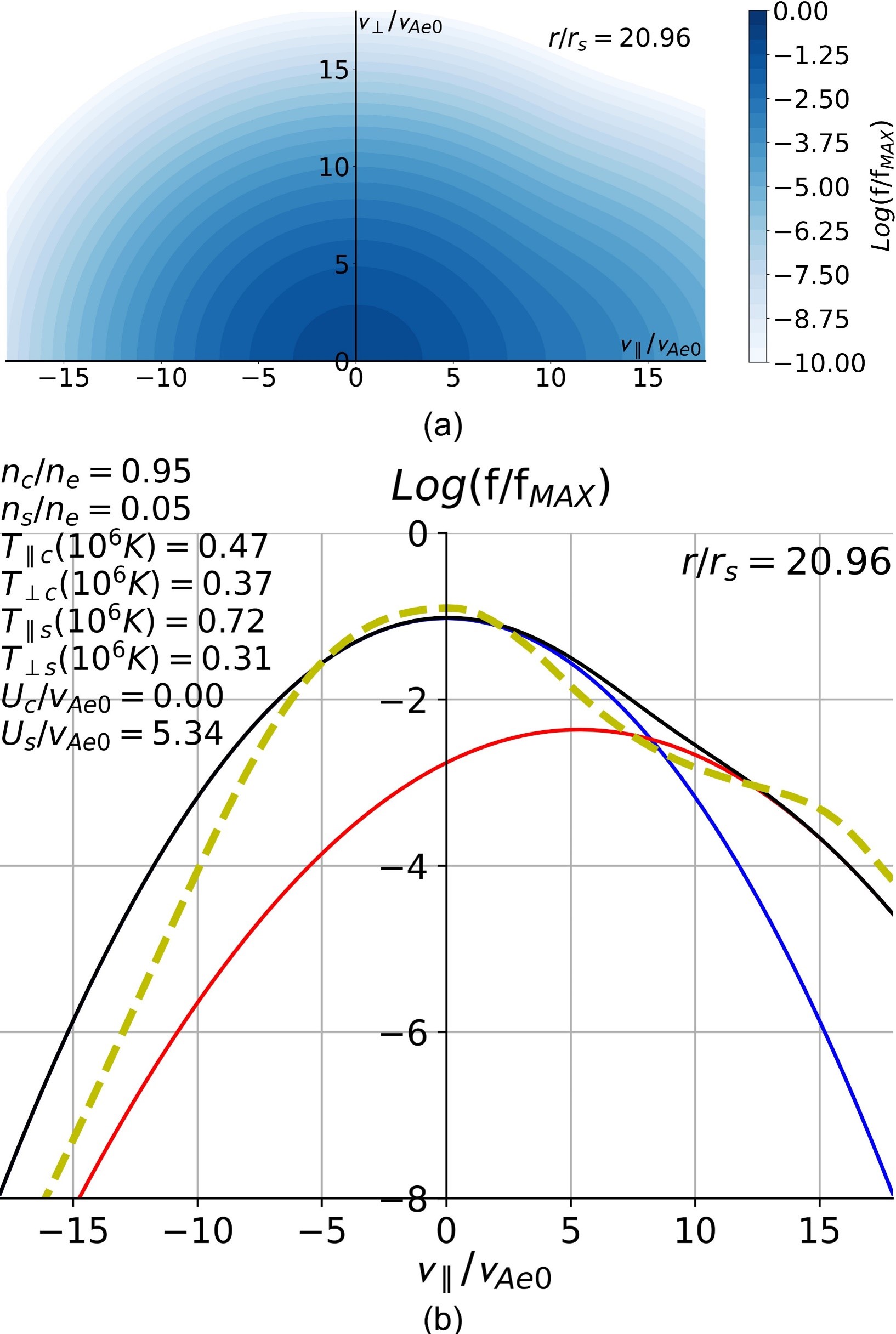}
  \caption{(a) The fitted electron VDF measured by PSP at a distance of $r/r_s=20.96$ on 27 September 2020 at 20:24:18 UTC during encounter 6; (b) the corresponding one-dimensional cut of the electron VDF in the $v_\parallel$-direction, with the fit parameters in the top-left corner. The blue and red solid curves are the fit results for the core and strahl based on the PSP data. The black solid curve is the sum of the core and strahl fits, and the yellow dashed curve is the same as the black dashed curve in Fig.~\ref{1Dresults}.}
  \label{PSP}
\end{figure}

Fig.~\ref{PSP}a shows, as a typical example, the fitted electron VDF measured by PSP at a distance of $r/r_s=20.96$ on 27 September 2020 at 20:24:18 UTC during encounter 6. Fig.~\ref{PSP}b is the corresponding one-dimensional cut of the electron VDF in the $v_\parallel$-direction. The blue and red solid curves are the fitted VDFs for the core and strahl from the PSP data. We provide the fit parameters in the top-left corner of Fig.~\ref{PSP}b. The black solid curve is the sum of the core and strahl fits, and the yellow dashed curve is the same as the black dashed curve in Fig.~\ref{1Dresults} from our numerical results. 

Comparing the PSP measurement with the fitted VDF  from our numerical results in Fig.~\ref{FittedVDF}, we find that our model  produces $n_s/n_e=7\%$ at $r/r_s=20$ which is slightly greater than the observed value. The values of $T_{\parallel s}$ and $T_{\perp s}$ from our model are 1.2 and 1.7 times greater, while the values of $T_{\parallel c}$ and $T_{\perp c}$ from our model are 0.7 and 0.9 times less than the values from the PSP observation. The shoulder-like strahl structure at around $v_\parallel/v_{Ae0}\gtrsim 10$ is more distinct in our model results than in the PSP data. We find a core temperature anisotropy with $T_{\perp c}<T_{\parallel c}$ in Fig.~\ref{PSP}b, which is opposite to the anisotropy found in Fig.~\ref{FittedVDF}b. We note that $U_c$ is zero in Fig.~\ref{PSP} while $U_c$ in Fig.~\ref{FittedVDF} is slightly negative; however, this difference is likely due to the choice of reference frame in the PSP level-3 data and associated uncertainties when $U_c$ is small. Lastly, our model produces $U_s/v_{Ae0}=5.59$ at $r/r_s=20$ which is close to the observed value.

\subsection{Oblique Fast-magnetosonic/Whistler Instability} \label{4.4}
We now investigate the possibility for the oblique FM/W instability to scatter strahl electrons into the halo as the solar wind expands into the heliosphere. The oblique FM/W instability has received major attention lately as a mechanism to explain the halo formation  \citep{2019ApJ...871L..29V,2019ApJ...886..136V,L_pez_2020,Jeong_2020,Micera_2020,Micera_2021,refId00,Sun_2021}. For this investigation, we compare our fit parameters from Section~\ref{4.2}  with the theoretically predicted threshold for the oblique FM/W instability in the low-$\beta_{\parallel c}$ regime given by \citet{2019ApJ...886..136V}. According to this framework, the oblique FM/W instability is unstable if
\begin{equation}\label{thres}
\begin{split}
U_s\gtrsim 3v_{\parallel th,c}.
\end{split}
\end{equation}
\begin{figure}[ht]
  \centering
  \includegraphics[width=\columnwidth]{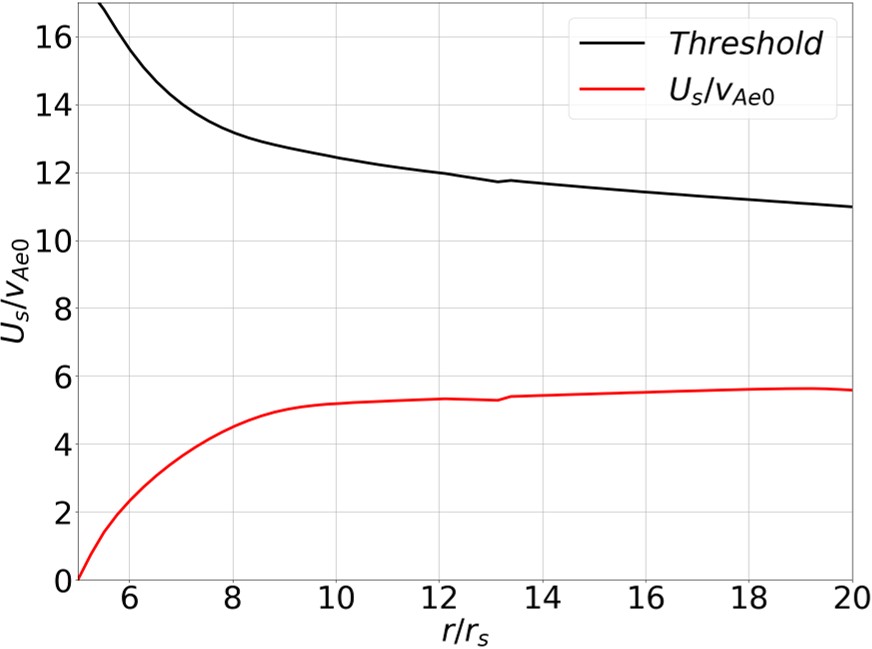}
  \caption{Comparison of the strahl bulk velocity with the threshold for the oblique FM/W instability. The red curve  corresponds to $U_s/v_{Ae0}$, and the black curve shows $3v_{\parallel th,c}/v_{Ae0}$. According to Eq.~(\ref{thres}), the oblique FM/W instability is unstable if $U_s\gtrsim 3v_{\parallel th,c}$. In our region of interest, the electron strahl does not cross the instability threshold.}
  \label{threshold}
\end{figure}

Fig.~\ref{threshold} shows $U_s/v_{Ae0}$ as a red solid curve and the threshold from Eq.~(\ref{thres}) normalized in units of $v_{Ae0}$ as a black solid curve, both as functions of radial distance. In our region of interest, the plasma does not cross the threshold for the oblique FM/W instability. This suggests that the electron strahl, under the typical parameters reproduced by our model, is not scattered by the oblique FM/W instability at these distances from the Sun. However, $U_s$ stays approximately constant at distances greater than $r/r_s=15$ while the threshold decreases with distance due to the decrease in $T_{\parallel c}$. The difference between the strahl speed and the threshold decreases with increasing distance. Therefore, assuming that this trend continues, we hypothesize that the electron strahl scattering by the oblique FM/W instability sets in at a critical distance $r_{\mathrm{crit}}>20r_s$  from the Sun. The value of $r_{\mathrm{crit}}$ is not known yet. In addition, the oblique FM/W instability transitions into a high-$\beta_{\parallel c}$ regime when $\beta_{\parallel c}\gtrsim 0.25$ \citep{2019ApJ...886..136V}, which occurs beyond $r/r_s=20$. We also note that, within the natural variability of the solar-wind parameters, crossings of the oblique FM/W-instability threshold can occur sporadically at times of particularly fast strahl or cold core conditions. These outlier conditions are not covered in our representative model.

\section{Discussion and Conclusions} \label{5}

In this paper, we derive a kinetic transport equation to describe the kinetic evolution of the solar-wind electrons in a Parker-spiral magnetic field. In the derivation of our kinetic transport equation, we work in a reference frame that rotates with the $r$-dependent Parker angle and  moves with the accelerating solar-wind bulk velocity. Based on Hamiltonian mechanics, we derive the external electromagnetic forces and complete our kinetic transport equation.

We evaluate an analytical model to explain the electron-strahl evolution in the very inner heliosphere, which is especially of great importance to understand the heat flux in the suprathermal part of the electron VDF. Therefore, we focus on the range of heliocentric distances between the corona, where collisions are more important, and distances up to $r/r_s=20$, which have not been fully sampled by in-situ spacecraft. To solve the three-dimensional kinetic transport equation (2D in velocity space and 1D in configuration space), we devise a numerical approach based on the combination of the Crank--Nicolson scheme and a forward Euler scheme. 

We show that the solar-wind electron VDF evolves through the combined effects of the accelerating solar wind, ballistic particle streaming, the parallel (to the magnetic field) electric field, the magnetic mirror force, the Parker-spiral geometry, and Coulomb collisions. Our kinetic evolution shows the formation of the electron strahl through the ballistic particle streaming and magnetic mirror force, a sunward shift of the electron core in velocity space through the electric field, and a decrease of the electron density in agreement with the fluid continuity equation for the electrons. These  results are in clear agreement with previous work on the  kinetic electron evolution \citep{1997JGR...102.4701L,2000JGR...105...35L,Vocks_2003,Smith_2012,Landi_2012,Tang_2020,https://doi.org/10.1029/2020JA028864}. 

We apply a fitting routine to our numerical results for the comparison of the electron core and strahl parameters with PSP observations. At  $r/r_s\sim 20$, the strahl density in our model is 7\% of the total electron density, compared to 5\% in typical observations from PSP \citep{Abraham2021_submitted}. The strahl parallel and perpendicular temperatures from our model are about  1.2 and 1.7 times greater, while the core parallel and perpendicular temperatures from our model are 0.7 and 0.9 times less than the observed temperatures. Our numerical model creates a slightly stronger strahl than observed, possibly because we set our lower-boundary VDF to a $\kappa$-distribution with $\kappa=8$ instead of a Maxwellian. We confirm that, if the lower-boundary VDF is a $\kappa$-distribution, the strahl formation is more distinct. We attribute this minor difference to the increased population of suprathermal strahl seed particles in the tail of the lower-boundary $\kappa$-distribution compared to a Maxwellian.

The core temperature anisotropy in our model ($T_{\perp c}>T_{\parallel c}$ in Fig.~\ref{FittedVDF}b) is opposite to the anisotropy in the PSP observation ($T_{\perp c}<T_{\parallel c}$ in Fig.~\ref{PSP}b). In our model results, the core anisotropy with $T_{\perp c}>T_{\parallel c}$ is likely generated because the sunward-streaming electrons in the VDF (i.e., those electrons with $U+v_{\parallel}\cos\theta<0$) de-focus with increasing time. A possible explanation for the difference in core anisotropy between our numerical results and  the observations is that, in reality, the core electrons already have a temperature anisotropy with $T_{\perp c}<T_{\parallel c}$ at the lower boundary of our integration domain. Additional core parallel-heating mechanisms, such as  electron Landau damping of kinetic Alfv\'en waves \citep{2019NatCo..10..740C}, are a possible explanation for this difference in core anisotropy. The strahl bulk velocity in our model increases between  $r/r_s=5$ and $r/r_s=15$. It  then stays constant with distance at a value of about $U_s/v_{Ae0}=5.6$, which is close to the typically observed value.   Lastly, by comparing our results with the threshold for the oblique FM/W instability in the low-$\beta_{\parallel c}$ regime, we find that the electron strahl is not scattered by this particular instability for typical solar-wind conditions in our region of interest.

Beyond $r/r_s=20$, we expect that the electron strahl continues to evolve because the electron temperature continuously decreases and the magnetic mirror force is still effective. However, with further distance, this type of strahl processing becomes ineffective due to the weakened gradients of the magnetic field and temperature \citep{https://doi.org/10.1029/2008JA013294,https://doi.org/10.1002/2015JA021368,Moncuquet_2020}. Then, the initially formed strahl traverses towards larger distances without undergoing a significant evolution (apart from a density decrease due to expansion), unless it experiences local scattering mechanisms. As suggested in Section~\ref{4.4}, if the strahl bulk velocity stays constant with distance, the threshold for the oblique FM/W instability crosses the strahl bulk velocity at a critical distance $r_{\mathrm{crit}}/r_s>20$ because $T_{\parallel c}$ continues to decrease with distance. However, the plasma transitions from the low-$\beta_{\parallel c}$ regime into the high-$\beta_{\parallel c}$ regime of the oblique FM/W instability with distance.  The transition between both regimes is defined at $\beta_{\parallel c}\approx 0.25$ \citep{2019ApJ...886..136V}, which is not crossed at the distances explored in our model (see Fig.~\ref{fittedplots}d). If the plasma does not fulfill Eq.~(\ref{thres}) before it reaches the high-$\beta_{\parallel c}$ regime, we must consider a greater threshold for the oblique FM/W instability as given by \citet{2019ApJ...886..136V}. The extension of our model to larger heliocentric distances and the comparison of the instability thresholds in these different regimes lie beyond the scope of this work.

Evidence for strahl scattering through the oblique FM/W instability in the near-Sun environment has been recently provided based on PSP data \citep{refId00}. In a small selection of cases, the observed electron strahl parameters  at heliocentric distances  $\lesssim 50r_s$ cross the threshold for the oblique FM/W instability. However, in agreement with our results, the majority of the parameter combinations are found to be stable with respect to this instability. This leaves the question open as to what process regulates the electron heat flux in the inner heliosphere. 

Our results suggest that the average strahl evolution within $r/r_s=20$ from the Sun is dominated by the kinetic effects included in our transport equation rather than wave--particle processes such as instabilities and resonant dissipation. However, wave--particle interactions are a possible explanation for the slight discrepancy between our model output and the PSP observations in terms of $T_{\parallel c}$.

\acknowledgments

D.V.~is supported by STFC Ernest Rutherford Fellowship ST/P003826/1. C. V. is supported by the Deutsche Forschungsgemeinschaft (DFG, German Research
Foundation) -- VO 2123/1-1. J. B. A. is supported by STFC grant ST/T506485/1. D.V., C.J.O., A.N.F., D.S., and L.B.~are supported by STFC Consolidated Grant ST/S000240/1. R. T. W. is funded by STFC consolidated Grant ST/V006320/1. J.A.A.R.~is supported by the ESA Networking/Partnering Initiative (NPI) contract 4000127929/19/NL/MH/mg and the Colombian programme Pasaporte a la Ciencia, Foco Sociedad -- Reto 3, ICETEX grant 3933061. M. B. is supported by a UCL Impact Studentship, joint funded by the ESA NPI contract 4000125082/18/NL/MH/ic. We appreciate helpful discussions with Christopher Chen. This work was discussed at the ``Joint Electron Project'' at MSSL.

\appendix
\section{Numerical Analysis of a 3-Dimensional Differential Equation}\label{A}
In our previous paper \citep{Jeong_2020}, we present a mathematical approach using a two-layer matrix to numerically solve a general two-dimensional differential equation based on the Crank--Nicolson scheme. In the present paper, we combine that approach with a forward Euler scheme to numerically solve our three-dimensional kinetic transport equation, Eq.~(\ref{convecff}).

For the sake of generality, we solve Eq.~(\ref{convecff}) with arbitrary coefficients for all terms:
\begin{equation}\label{simf}
\begin{split}
\frac{\partial  f_e }{\partial  t}=\alpha f_e+\alpha^{r}\frac{\partial f_e}{\partial r}+\alpha^{\perp}\frac{\partial f_e}{\partial v_\perp}+\alpha^{\parallel}\frac{\partial f_e}{\partial v_\parallel}+\alpha^{\perp \perp}\frac{\partial^2 f_e}{\partial v_\perp^2}+\alpha^{\parallel \parallel}\frac{\partial^2 f_e}{\partial v_\parallel^2}+\alpha^{\perp \parallel}\frac{\partial^2 f_e}{\partial v_\perp v_\parallel},
\end{split}
\end{equation}
where $\alpha, \alpha^{r}, \alpha^{\perp}, \alpha^{\parallel}, \alpha^{\perp \perp}, \alpha^{\parallel \parallel}$ and $\alpha^{\perp \parallel}$ explicitly depend on $r, v_\perp$, and $v_\parallel$. We divide velocity space into $N_v\times N_v$ steps with equal step sizes of $\Delta v$ by defining the outer boundaries of velocity space as $\pm v_{\perp}^{\max}$ and $\pm v_{\parallel}^{\max}$. In addition, we divide the radial space into $N_r$ steps with equal step sizes of $\Delta r$ by defining the inner and outer boundaries of radial space as $r^{\mathrm{in}}$ and $r^{\mathrm{out}}$. The $v_\perp$-index $M$ and the $v_\parallel$-index $N$ both step through 1, 2, \dots, $N_v$, and the $r$-index $L$ steps through 1, 2, \dots, $N_r$. We define the discrete velocity and radial coordinates as $v_{\perp M} \equiv -v_{\perp}^{\max}+(M-1)\Delta v$, $v_{\parallel N} \equiv -v_{\parallel}^{\max}+(N-1)\Delta v$ and $r_{L} \equiv r^{\mathrm{in}}+(L-1)\Delta r$. We note that this definition introduces negative $v_{\perp}$-values that only serve numerical purposes, and we neglect them in our computational results. We also divide the time $t$ into equal step sizes $\Delta t$, and the $t$-index $T$ steps through $1, 2, \cdots$. We define the discrete time as $t^T \equiv (T-1)\Delta t$. We then define the discrete VDF as $f_{L,M,N}^{T}\equiv f_e(r_L, \mathit{v}_{\perp M},\mathit{v}_{\parallel N}, t^T )$. 

For the discretization of the velocity and radial derivatives, we adopt the two-point central difference operator. We then apply the Crank--Nicolson scheme to the velocity derivatives and the source term, and a forward Euler scheme to the radial derivative in Eq.~(\ref{simf}), which leads to
\begin{equation}\label{dsimf}
\begin{split}
&\frac{f^{T+1}_{L,M,N}\!-\!f^T_{L,M,N}}{ \Delta t}\!=\!\frac{\alpha_{L,M,N}}{2}\!\left[ f^{T+1}_{L,M,N}\!+\!f^{T}_{L,M,N} \right]\!+\!\alpha^{r}_{L,M,N}\!\left(\frac{\partial f_e}{\partial r}\right)^{T}_{L,M,N}\!+\!\frac{\alpha^{\perp}_{L,M,N}}{2}\!\left[\left(\frac{\partial f_e}{\partial v_\perp}\right)^{T+1}_{L,M,N}\!+\!\left(\frac{\partial f_e}{\partial v_\perp}\right)^{T}_{L,M,N}  \right]\\
&+\frac{\alpha^{\parallel}_{L,M,N}}{2}\left[\left(\frac{\partial f_e}{\partial v_\parallel}\right)^{T+1}_{L,M,N}+\left(\frac{\partial f_e}{\partial v_\parallel}\right)^{T}_{L,M,N}  \right]+\frac{\alpha^{\perp \perp}_{L,M,N}}{2}\left[\left(\frac{\partial^2 f_e}{\partial v_\perp^2}\right)^{T+1}_{L,M,N}+\left(\frac{\partial^2 f_e}{\partial v_\perp^2}\right)^{T}_{L,M,N}  \right]\\
&+\frac{\alpha^{\parallel \parallel}_{L,M,N}}{2}\left[\left(\frac{\partial^2 f_e}{\partial v_\parallel^2}\right)^{T+1}_{L,M,N}+\left(\frac{\partial^2 f_e}{\partial v_\parallel^2}\right)^{T}_{L,M,N}  \right]+\frac{\alpha^{\perp \parallel}_{L,M,N}}{2}\left[\left(\frac{\partial^2 f_e}{\partial v_\perp \partial v_\parallel}\right)^{T+1}_{L,M,N}+\left(\frac{\partial^2 f_e}{\partial v_\perp \partial v_\parallel}\right)^{T}_{L,M,N}  \right],
\end{split}
\end{equation}
where
\begin{equation}
\begin{split}
\left(\frac{\partial f_e}{\partial r}\right)^{T}_{L,M,N}=\frac{f^T_{L+1,M,N}-f^T_{L-1,M,N}}{2 \Delta r},
\end{split}
\end{equation}
\begin{equation}
\begin{split}
\left(\frac{\partial f_e}{\partial v_\perp}\right)^{T}_{L,M,N}=\frac{f^T_{L,M+1,N}-f^T_{L,M-1,N}}{2 \Delta v},
\end{split}
\end{equation}
\begin{equation}
\begin{split}
\left(\frac{\partial f_e}{\partial v_\parallel}\right)^{T}_{L,M,N}=\frac{f^T_{L,M,N+1}-f^T_{L,M,N-1}}{2 \Delta v},
\end{split}
\end{equation}
\begin{equation}
\begin{split}
&\left(\frac{\partial^2 f_e}{\partial v_\perp^2}\right)^{T}_{L,M,N}=\frac{f^T_{L,M+2,N}-2f^T_{L,M,N}+f^T_{L,M-2,N}}{4 (\Delta v)^2},
\end{split}
\end{equation}
\begin{equation}
\begin{split}
\left(\frac{\partial^2 f_e}{\partial v_\parallel^2}\right)^{T}_{L,M,N}=\frac{f^T_{L,M,N+2}-2f^T_{L,M,N}+f^T_{L,M,N-2}}{4 (\Delta v)^2},
\end{split}
\end{equation}
\begin{equation}
\begin{split}
\left(\frac{\partial^2 f_e}{\partial v_\perp \partial v_\parallel}\right)^{T}_{L,M,N}=\frac{f^T_{L,M+1,N+1}-f^T_{L,M+1,N-1}-f^T_{L,M-1,N+1}+f^T_{L,M-1,N-1}}{4 (\Delta v)^2},
\end{split}
\end{equation}
and $\alpha_{L,M,N}$, $\alpha^{r}_{L,M,N}$, $\alpha^{\perp}_{L,M,N}$, $\alpha^{\parallel}_{L,M,N}$, $\alpha^{\perp \perp}_{L,M,N}$, $\alpha^{\parallel \parallel}_{L,M,N}$, and $\alpha^{\perp \parallel}_{L,M,N}$ are the values of $\alpha, \alpha^{r}, \alpha^{\perp}, \alpha^{\parallel}, \alpha^{\perp \perp}, \alpha^{\parallel \parallel}$ and $\alpha^{\perp \parallel}$, estimated at $r=r_L$, $v_\perp=v_{\perp M}$, and $v_\parallel=v_{\parallel N}$.  Rearranging  Eq.~(\ref{dsimf}) yields
\begin{equation}\label{resolvedCN}
\begin{split}
&-\frac{\mu_{vv}}{8}\alpha^{\perp \perp}_{L,M,N}f_{L,M-2,N}^{T+1}-\frac{\mu_{vv}}{8}\alpha^{\perp \parallel}_{L,M,N}f_{L,M-1,N-1}^{T+1}+\frac{\mu_{v}}{4}\alpha^{\perp}_{L,M,N}f_{L,M-1,N}^{T+1}+\frac{\mu_{vv}}{8}\alpha^{\perp \parallel}_{L,M,N}f_{L,M-1,N+1}^{T+1}\\
&-\frac{\mu_{vv}}{8}\alpha^{\parallel \parallel}_{L,M,N}f_{L,M,N-2}^{T+1}+\frac{\mu_{v}}{4}\alpha^{\parallel}_{L,M,N}f_{L,M,N-1}^{T+1}+\left ( 1-\frac{\Delta t}{2} \alpha_{L,M,N}+\frac{\mu_{vv}}{4}\alpha^{\perp \perp}_{L,M,N}+\frac{\mu_{vv}}{4}\alpha^{\parallel \parallel}_{L,M,N} \right)f_{L,M,N}^{T+1}\\
&-\frac{\mu_{v}}{4}\alpha^{\parallel}_{L,M,N}f_{L,M,N+1}^{T+1}-\frac{\mu_{vv}}{8}\alpha^{\parallel \parallel}_{L,M,N}f_{L,M,N+2}^{T+1}+\frac{\mu_{vv}}{8}\alpha^{\perp \parallel}_{L,M,N}f_{L,M+1,N-1}^{T+1}-\frac{\mu_{v}}{4}\alpha^{\perp}_{L,M,N}f_{L,M+1,N}^{T+1}\\
&-\frac{\mu_{vv}}{8}\alpha^{\perp \parallel}_{L,M,N}f_{L,M+1,N+1}^{T+1}-\frac{\mu_{vv}}{8}\alpha^{\perp \perp}_{L,M,N}f_{L,M+2,N}^{T+1}\\
&=\\
&\frac{\mu_{vv}}{8}\alpha^{\perp \perp}_{L,M,N}f_{L,M-2,N}^{T}+\frac{\mu_{vv}}{8}\alpha^{\perp \parallel}_{L,M,N}f_{L,M-1,N-1}^{T}-\frac{\mu_{v}}{4}\alpha^{\perp}_{L,M,N}f_{L,M-1,N}^{T}-\frac{\mu_{vv}}{8}\alpha^{\perp \parallel}_{L,M,N}f_{L,M-1,N+1}^{T}\\
&+\frac{\mu_{vv}}{8}\alpha^{\parallel \parallel}_{L,M,N}f_{L,M,N-2}^{T}-\frac{\mu_{v}}{4}\alpha^{\parallel}_{L,M,N}f_{L,M,N-1}^{T}+\left ( 1+\frac{\Delta t}{2} \alpha_{L,M,N}-\frac{\mu_{vv}}{4}\alpha^{\perp \perp}_{L,M,N}-\frac{\mu_{vv}}{4}\alpha^{\parallel \parallel}_{L,M,N} \right)f_{L,M,N}^{T}\\
&+\frac{\mu_{v}}{4}\alpha^{\parallel}_{L,M,N}f_{L,M,N+1}^{T}+\frac{\mu_{vv}}{8}\alpha^{\parallel \parallel}_{L,M,N}f_{L,M,N+2}^{T}-\frac{\mu_{vv}}{8}\alpha^{\perp \parallel}_{L,M,N}f_{L,M+1,N-1}^{T}+\frac{\mu_{v}}{4}\alpha^{\perp}_{L,M,N}f_{L,M+1,N}^{T}\\
&+\frac{\mu_{vv}}{8}\alpha^{\perp \parallel}_{L,M,N}f_{L,M+1,N+1}^{T}+\frac{\mu_{vv}}{8}\alpha^{\perp \perp}_{L,M,N}f_{L,M+2,N}^{T}+\frac{\mu_r}{2}\alpha^{r}_{L,M,N}\left[f_{L+1,M,N}^{T}-f_{L-1,M,N}^{T}\right]
\end{split}
\end{equation}
where $\mu_r=\Delta t/\Delta r$, $\mu_v=\Delta t/\Delta v$, and $\mu_{vv}=\Delta t/\Delta v^2$. On both sides of Eq.~(\ref{resolvedCN}), we group the terms by the same $v_\perp$-index of VDFs with $r$-index of $L$ and arrange these groups in increasing order in $v_\perp$-index of VDFs. In each group, we arrange terms in increasing order in $v_\parallel$-index of VDFs. On the right-hand side of Eq.~(\ref{resolvedCN}), we leave VDFs with $r$-index of $L+1$ and $L-1$ in the form of Euler scheme. Eq.~(\ref{resolvedCN}) is a three-dimensional set of algebraic matrix equations. Eq.~(\ref{resolvedCN}) is implicit for the two-dimensional velocity space, resulting from the Crank--Nicolson scheme, and explicit for the one-dimensional configuration ($r$) space, resulting from  the Euler scheme. The arrangement shown in Eq.~(\ref{resolvedCN}) allows us to transform Eq.~(\ref{resolvedCN}) into a one-dimensional explicit equation in $r$-space by applying the concept of a two-layer matrix to the velocity space \citep{Jeong_2020}. 


We transform all terms with $r$-index $L$ and $v_{\perp}$-index $M$ in the VDF on both sides of Eq.~(\ref{resolvedCN}) into the tridiagonal matrices $\mathbb A_{L,M}^{(1)} \mathbb F_{L,M}^{T+1}$ and $\mathbb A_{L,M}^{(2)}\mathbb F_{L,M}^{T}$, where $\mathbb F_{L,M}^{T}\equiv[
f_{L,M,1}^{T} \: f_{L,M,2}^{T} \: \cdots \: f_{L,M,N_v}^{T}]^\textbf T_{ 1 \times  N_v}$ and $\textbf T$ represents the transpose of a matrix. We then find that
\begin{equation}
\mathbb A_{L,M}^{(1)} \!\equiv\!
\begin{bmatrix}
\alpha_{L,M,1}^{(1)} & -\frac{\mu_v}{4}\alpha^{\parallel}_{L,M,1} & -\frac{\mu_{vv}}{8}\alpha^{\parallel \parallel}_{L,M,1} & 0 & 0 & \cdots & 0\\
\frac{\mu_v}{4}\alpha^{\parallel}_{L,M,2} & \alpha_{L,M,2}^{(1)} & -\frac{\mu_v}{4}\alpha^{\parallel}_{L,M,2} & -\frac{\mu_{vv}}{8}\alpha^{\parallel \parallel}_{L,M,2} & 0 & \cdots & 0\\
-\frac{\mu_{vv}}{8}\alpha^{\parallel \parallel}_{L,M,3} & \frac{\mu_v}{4}\alpha^{\parallel}_{L,M,3} & \alpha_{L,M,3}^{(1)} & -\frac{\mu_v}{4}\alpha^{\parallel}_{L,M,3} & -\frac{\mu_{vv}}{8}\alpha^{\parallel \parallel}_{L,M,3} & \cdots & 0 \\
 & \vdots &  & \ddots &  & \vdots &  \\
0 & \cdots & 0 & 0 & -\frac{\mu_{vv}}{8}\alpha^{\parallel \parallel}_{L,M,N_v} & \frac{\mu_v}{4}\alpha^{\parallel}_{L,M,N_v} & \alpha_{L,M,N_v}^{(1)} \\
\end{bmatrix}
_{N_v \times N_v},
\end{equation}
\begin{equation}
\begin{split}
\mathbf{\alpha}_{L,M,N}^{(1)}=1-\frac{\Delta t}{2}\alpha_{L,M,N}+\frac{\mu_{vv}}{4}\alpha^{\perp \perp}_{L,M,N}+\frac{\mu_{vv}}{4}\alpha^{\parallel \parallel}_{L,M,N},
\end{split}
\end{equation}
\begin{equation}
\mathbb A_{L,M}^{(2)} \!\equiv\!
\begin{bmatrix}
\alpha_{L,M,1}^{(2)} & \frac{\mu_v}{4}\alpha^{\parallel}_{L,M,1} & \frac{\mu_{vv}}{8}\alpha^{\parallel \parallel}_{L,M,1} & 0 & 0 & \cdots & 0\\
-\frac{\mu_v}{4}\alpha^{\parallel}_{L,M,2} & \alpha_{L,M,2}^{(2)} & \frac{\mu_v}{4}\alpha^{\parallel}_{L,M,2} & \frac{\mu_{vv}}{8}\alpha^{\parallel \parallel}_{L,M,2} & 0 & \cdots & 0\\
\frac{\mu_{vv}}{8}\alpha^{\parallel \parallel}_{L,M,3} & -\frac{\mu_v}{4}\alpha^{\parallel}_{L,M,3} & \alpha_{L,M,3}^{(2)} & \frac{\mu_v}{4}\alpha^{\parallel}_{L,M,3} & \frac{\mu_{vv}}{8}\alpha^{\parallel \parallel}_{L,M,3} & \cdots & 0 \\
 & \vdots &  & \ddots &  & \vdots &  \\
0 & \cdots & 0 & 0 & \frac{\mu_{vv}}{8}\alpha^{\parallel \parallel}_{L,M,N_v} & -\frac{\mu_v}{4}\alpha^{\parallel}_{L,M,N_v} & \alpha_{L,M,N_v}^{(2)} \\
\end{bmatrix}
_{N_v \times N_v},
\end{equation}
and
\begin{equation}
\begin{split}
\alpha_{L,M,N}^{(2)}=1+\frac{\Delta t}{2}\alpha_{L,M,N}-\frac{\mu_{vv}}{4}\alpha^{\perp \perp}_{L,M,N}-\frac{\mu_{vv}}{4}\alpha^{\parallel \parallel}_{L,M,N}.
\end{split}
\end{equation}
We transform all terms with $r$-index  $L$ and $v_{\perp}$-index  $M-1$ in the VDF on both sides of Eq.~(\ref{resolvedCN}) into the tridiagonal matrices $\mathbb B_{L,M} \mathbb F_{L,M-1}^{T+1}$ and $-\mathbb B_{L,M} \mathbb F_{L,M-1}^{T}$,
where
\begin{equation}
\mathbb B_{L,M} \equiv
\begin{bmatrix}
\frac{\mu_v}{4}\alpha^{\perp}_{L,M,1} & \frac{\mu_{vv}}{8}\alpha^{\perp \parallel}_{L,M,1} & 0 & \cdots & 0\\
-\frac{\mu_{vv}}{8}\alpha^{\perp \parallel}_{L,M,2} & \frac{\mu_v}{4}\alpha^{\perp}_{L,M,2} & \frac{\mu_{vv}}{8}\alpha^{\perp \parallel}_{L,M,2}  & \cdots & 0 \\
 & \vdots & \ddots &  & \vdots \\
0 & \cdots  & 0 & -\frac{\mu_{vv}}{8}\alpha^{\perp \parallel}_{L,M,N_v} & \frac{\mu_v}{4}\alpha^{\perp}_{L,M,N_v} \\
\end{bmatrix}
_{N_v \times N_v}.
\end{equation}
We transform all terms with $r$-index  $L$ and $v_{\perp}$-index  $M+1$ in the VDF on both sides of Eq.~(\ref{resolvedCN}) into the tridiagonal matrices $-\mathbb B_{L,M} \mathbb F_{L,M+1}^{T+1}$ and $\mathbb B_{L,M} \mathbb F_{L,M+1}^{T}$. Likewise, we transform all terms with $r$-index  $L$ and $v_{\perp}$-index  $M-2$ in the VDF on both sides of Eq.~(\ref{resolvedCN}) into the tridiagonal matrices $-\mathbb C_{L,M} \mathbb F_{L,M-2}^{T+1}$ and $\mathbb C_{L,M} \mathbb F_{L,M-2}^{T}$, where
\begin{equation}
\mathbb C_{L,M} \equiv
\begin{bmatrix}
\frac{\mu_{vv}}{8}\alpha^{\perp \perp}_{L,M,1} & 0  & \cdots & 0\\
0 & \frac{\mu_{vv}}{8}\alpha^{\perp \perp}_{L,M,2}  & \cdots & 0\\
 \vdots &  & \ddots & \vdots \\
0 & \cdots & 0  & \frac{\mu_{vv}}{8}\alpha^{\perp \perp}_{L,M,N_v}\\
\end{bmatrix}
_{N_v \times N_v}.
\end{equation}
We transform all terms with $r$-index  $L$ and $v_{\perp}$-index  $M+2$ in the VDF on both sides of Eq.~(\ref{resolvedCN}) into the tridiagonal matrices $-\mathbb C_{L,M} \mathbb F_{L,M+2}^{T+1}$ and $\mathbb C_{L,M} \mathbb F_{L,M+2}^{T}$. Likewise, we transform all terms with $r$-index  $L-1$ and $v_{\perp}$-index  $M$ in the VDF on the right-hand side of Eq.~(\ref{resolvedCN}) into the tridiagonal matrices $-\mathbb D_{L,M} \mathbb F_{L-1,M}^{T}$, where
\begin{equation}
\mathbb D_{L,M} \equiv
\begin{bmatrix}
\frac{\mu_{r}}{2}\alpha^{r}_{L,M,1} & 0  & \cdots & 0\\
0 & \frac{\mu_{r}}{2}\alpha^{r}_{L,M,2}  & \cdots & 0\\
 \vdots &  & \ddots & \vdots \\
0 & \cdots & 0  & \frac{\mu_{r}}{2}\alpha^{r}_{L,M,N_v}\\
\end{bmatrix}
_{N_v \times N_v}.
\end{equation}
Lastly, we transform all terms with $r$-index  $L+1$ and $v_{\perp}$-index  $M$ in the VDF on the right-hand side of Eq.~(\ref{resolvedCN}) into the tridiagonal matrices $\mathbb D_{L,M} \mathbb F_{L+1,M}^{T}$. By combining all transformed matrices, Eq.~(\ref{resolvedCN}) becomes a two-dimensional set of algebraic matrix equations given as
\begin{equation}\label{twoal}
\begin{split}
&-\! \mathbb C_{L,M} \mathbb F_{L,M-2}^{T+1}\!+\!\mathbb B_{L,M} \mathbb F_{L,M-1}^{T+1}\!+\!\mathbb A_{L,M}^{(1)} \mathbb F_{L,M}^{T+1}\!-\!\mathbb B_{L,M} \mathbb F_{L,M+1}^{T+1}\!-\!\mathbb C_{L,M} \mathbb F_{L,M+2}^{T+1}\\
&=\mathbb C_{L,M} \mathbb F_{L,M-2}^{T}\!-\!\mathbb B_{L,M} \mathbb F_{L,M-1}^{T}\!+\!\mathbb A_{L,M}^{(2)} \mathbb F_{L,M}^{T}\!+\!\mathbb B_{L,M} \mathbb F_{L,M+1}^{T}\!+\!\mathbb C_{L,M} \mathbb F_{L,M+2}^{T}\!+\!\mathbb D_{L,M} \left[\mathbb F_{L+1,M}^{T}-\mathbb F_{L-1,M}^{T}\right].
\end{split}
\end{equation}
Eq.~(\ref{twoal}) describes the VDF evolution in  $r$- and $v_{\perp}$-space. However, each matrix term itself includes the VDF evolution in  $v_{\parallel}$-space. Once again, we transform all terms with $r$-index  $L$ in the VDF on both sides of Eq.~(\ref{twoal}) into the tridiagonal matrices $\mathbb{\mathbf A}_{L}^{(1)} \mathbb {\mathbf F}_{L}^{T+1}$ and $\mathbb {\mathbf A}_{L}^{(2)} \mathbb {\mathbf F}_{L}^{T}$, where $\mathbb {\mathbf F}_{L}^{T}\equiv[
\mathbb F_{L,1}^{T} \: \mathbb F_{L,2}^{T} \: \cdots \: \mathbb F_{L,N_v}^{T}]^\mathbf T_{ 1 \times  (N_v)^2}$,
\begin{equation}\label{Amatrix}
\mathbb {\mathbf A}_{L}^{(1)} \!\equiv\!
\begin{bmatrix}
\mathbb A_{L,1}^{(1)} & -\mathbb B_{L,1} & -\mathbb C_{L,1} & 0 & 0 & \cdots & 0\\
\mathbb B_{L,2} & \mathbb A_{L,2}^{(1)} & -\mathbb B_{L,2} & -\mathbb C_{L,2} & 0 & \cdots & 0\\
-\mathbb C_{L,3} & \mathbb B_{L,3} & \mathbb A_{L,3}^{(1)} & -\mathbb B_{L,3} & -\mathbb C_{L,3} & \cdots & 0 \\
 & \vdots &  & \ddots &  & \vdots &  \\
0 & \cdots & 0 & 0 & -\mathbb C_{L,N_v} & \mathbb B_{L,N_v} & \mathbb A_{L,N_v}^{(1)} \\
\end{bmatrix}
_{(N_v)^2 \times (N_v)^2},
\end{equation}
and
\begin{equation}
\mathbb {\mathbf A}_{L}^{(2)} \!\equiv\!
\begin{bmatrix}
\mathbb A_{L,1}^{(2)} & \mathbb B_{L,1} & \mathbb C_{L,1} & 0 & 0 & \cdots & 0\\
-\mathbb B_{L,2} & \mathbb A_{L,2}^{(2)} & \mathbb B_{L,2} & \mathbb C_{L,2} & 0 & \cdots & 0\\
\mathbb C_{L,3} & -\mathbb B_{L,3} & \mathbb A_{L,3}^{(2)} & \mathbb B_{L,3} & \mathbb C_{L,3} & \cdots & 0 \\
 & \vdots &  & \ddots &  & \vdots &  \\
0 & \cdots & 0 & 0 & \mathbb C_{L,N_v} & -\mathbb B_{L,N_v} & \mathbb A_{L,N_v}^{(2)} \\
\end{bmatrix}
_{(N_v)^2 \times (N_v)^2}.
\end{equation}
We transform the term with $r$-index $L-1$ in the VDF on the right-hand side of Eq.~(\ref{twoal}) into the tridiagonal matrices $-\mathbb {\mathbf D}_{L} \mathbb {\mathbf F}_{L-1}^{T}$, where
\begin{equation}
\mathbb{\mathbf D}_{L} \equiv
\begin{bmatrix}
\mathbb D_{L,1} & 0  & \cdots & 0\\
0 & \mathbb D_{L,2}  & \cdots & 0\\
 \vdots &  & \ddots & \vdots \\
0 & \cdots & 0  & \mathbb D_{L,N_v}\\
\end{bmatrix}
_{(N_v)^2 \times (N_v)^2}.
\end{equation}
Lastly, we transform the term with $r$-index  $L+1$ in the VDF on the right-hand side of Eq.~(\ref{twoal}) into the tridiagonal matrices $\mathbb {\mathbf D}_{L} \mathbb {\mathbf F}_{L+1}^{T}$. By combining all transformed matrices, Eq.~(\ref{twoal}) becomes a one-dimensional set of algebraic matrix equations given as
\begin{equation}\label{sol}
\begin{split}
\mathbb{\mathbf A}_{L}^{(1)} \mathbb{\mathbf  F}_{L}^{T+1}=\mathbb{\mathbf A}_{L}^{(2)} \mathbb{\mathbf F}_{L}^{T}+\mathbb{\mathbf D}_{L} \left[ \mathbb{\mathbf F}_{L+1}^{T}-\mathbb{\mathbf F}_{L-1}^{T}\right].
\end{split}
\end{equation}

Eq.~(\ref{sol}) is a one-dimensional explicit equation as a form of Euler scheme, and describes the VDF evolution in  $r$-space. However, each term in Eq.~(\ref{sol}) is in the form of a two-layer matrix. The outer matrices evolve $f_{L,M,N}^{T}$ in  $v_\perp$-space, and the inner matrices evolve $f_{L,M,N}^{T}$ in  $v_\parallel$-space during each time step. By multiplying Eq.~(\ref{sol}) with the inverse of the tridiagonal matrix $\mathbb{\mathbf A}_{L}^{(1)}$ from Eq.~(\ref{Amatrix}), $\mathbb{\mathbf F}_{L}^{T}$ evolves in the $r$-, $v_{\perp}$- and $v_{\parallel}$-spaces for one time step. Eq.~(\ref{sol}) is then repeated to step through further time steps as required.

\section{Smoothing Effect in  Velocity Space}\label{b}
\begin{figure}[ht]
  \centering
  \includegraphics[width=\columnwidth]{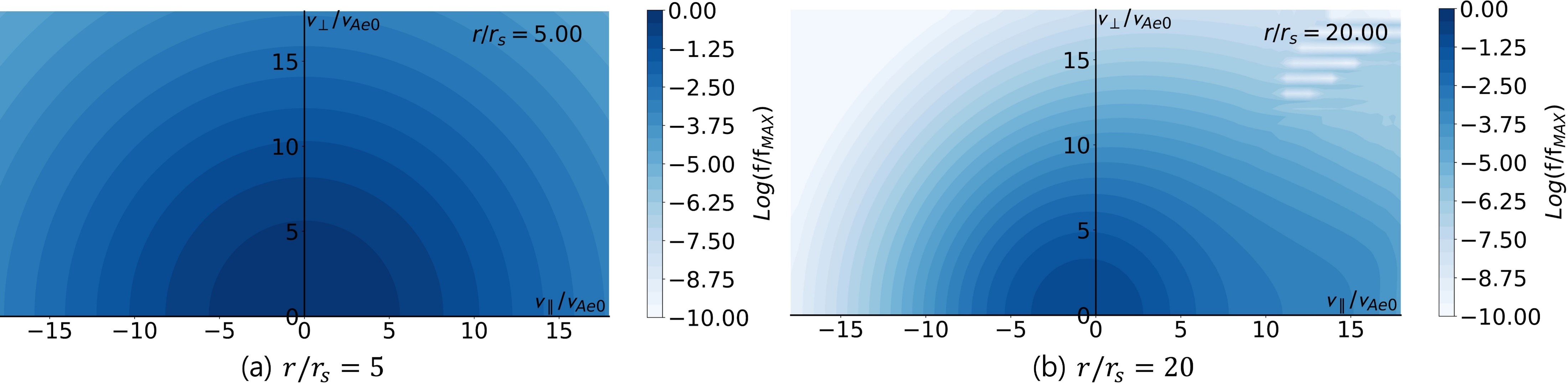}
  \caption{Numerical results without averaging through Eq.~(\ref{velocitysmooth}) in velocity space. Due to the limited velocity resolution, the numerical errors occur on the top-right corner in Fig.~\ref{2Dresults_without_smoothing}b. An animation of this figure is available. The animation shows the kinetic evolution of the electron VDF from $r/r_s=5$ to $r/r_s=20$, which is not averaged by Eq.~(\ref{velocitysmooth}). Panels (a) and (b) show the initial and final snapshot of the animation.}
  \label{2Dresults_without_smoothing}
\end{figure}

Fig.~\ref{2Dresults_without_smoothing} shows the results for the kinetic evolution of the electron VDF calculated with our numerical treatment described in Section~\ref{3}. However, the result shown in Fig.~\ref{2Dresults_without_smoothing} does not include the smoothing defined in Eq.~(\ref{velocitysmooth}), while Fig.~\ref{2Dresults} does.
In the top-right corner of Fig.~\ref{2Dresults_without_smoothing}b, numerical errors occur at very low values of the VDF. Comparing Fig.~\ref{2Dresults_without_smoothing} with Fig.~\ref{2Dresults}, our smoothing scheme, Eq.~(\ref{velocitysmooth}), only cleans out these numerical errors without changing other parts of the electron VDF.
 Using the fitting scheme from Section~\ref{4.2}, the fit parameters for the VDF shown in  Fig.~\ref{2Dresults_without_smoothing}b are $n_c/n_e=0.93$, $n_s/n_e=0.07$, $T_{\parallel c}=0.31\times 10^6\,K$, $T_{\perp c}(10^6K)=0.35$, $T_{\parallel s}=0.89\times 10^6\,K$, $T_{\perp s}=0.54\times 10^6\,K$, $U_c/v_{Ae0}=-0.18$ and $U_s/v_{Ae0}=5.59$. 
 These fit parameters are identical to the fit parameters from Fig.~\ref{2Dresults}b (see Fig.~\ref{FittedVDF}b). These errors are due to the limited velocity resolution and grow  over time.
 
 We conclude that Eq.~(\ref{velocitysmooth}) is an appropriate method to improve the numerical stability of our algorithm without affecting the physics captured in our model.

\bibliography{sample63}{}

\begin{thebibliography}{}
\expandafter\ifx\csname natexlab\endcsname\relax\def\natexlab#1{#1}\fi
\providecommand{\url}[1]{\href{#1}{#1}}
\providecommand{\dodoi}[1]{doi:~\href{http://doi.org/#1}{\nolinkurl{#1}}}
\providecommand{\doeprint}[1]{\href{http://ascl.net/#1}{\nolinkurl{http://ascl.net/#1}}}
\providecommand{\doarXiv}[1]{\href{https://arxiv.org/abs/#1}{\nolinkurl{https://arxiv.org/abs/#1}}}

\bibitem[{{Abraham} {et~al.}(2021){Abraham}, {Owen}, {Verscharen}, {Bakrania},
  {Stansby}, {Wicks}, {Nicolaou}, { Agudelo Rueda}, {Jeong}, {Whittlesey}, \&
  {Ber\v{c}i\v{c}}}]{Abraham2021_submitted}
{Abraham}, J.~B., {Owen}, C.~J., {Verscharen}, D., {et~al.} 2021, \apj

\bibitem[{{Badman} {et~al.}(2021){Badman}, {Bale}, {Rouillard}, {Bowen},
  {Bonnell}, {Goetz}, {Harvey}, {MacDowall}, {Malaspina}, \&
  {Pulupa}}]{2021A&A...650A..18B}
{Badman}, S.~T., {Bale}, S.~D., {Rouillard}, A.~P., {et~al.} 2021, \aap, 650,
  A18, \dodoi{10.1051/0004-6361/202039407}

\bibitem[{Bemporad(2017)}]{Bemporad_2017}
Bemporad, A. 2017, The Astrophysical Journal, 846, 86,
  \dodoi{10.3847/1538-4357/aa7de4}

\bibitem[{Berčič {et~al.}(2021)Berčič, Landi, \&
  Maksimović}]{https://doi.org/10.1029/2020JA028864}
Berčič, L., Landi, S., \& Maksimović, M. 2021, Journal of Geophysical
  Research: Space Physics, 126, e2020JA028864,
  \dodoi{https://doi.org/10.1029/2020JA028864}

\bibitem[{Boldyrev {et~al.}(2020)Boldyrev, Forest, \& Egedal}]{Boldyrev9232}
Boldyrev, S., Forest, C., \& Egedal, J. 2020, Proceedings of the National
  Academy of Sciences, 117, 9232, \dodoi{10.1073/pnas.1917905117}

\bibitem[{{Boldyrev} \& {Horaites}(2019)}]{2019MNRAS.489.3412B}
{Boldyrev}, S., \& {Horaites}, K. 2019, \mnras, 489, 3412,
  \dodoi{10.1093/mnras/stz2378}

\bibitem[{{Chen} {et~al.}(2019){Chen}, {Klein}, \&
  {Howes}}]{2019NatCo..10..740C}
{Chen}, C.~H.~K., {Klein}, K.~G., \& {Howes}, G.~G. 2019, Nature
  Communications, 10, 740, \dodoi{10.1038/s41467-019-08435-3}

\bibitem[{Chen {et~al.}(1972)Chen, Lai, Lin, \&
  Lin}]{https://doi.org/10.1029/JA077i001p00001}
Chen, W.~M., Lai, C.~S., Lin, H.~E., \& Lin, W.~C. 1972, Journal of Geophysical
  Research (1896-1977), 77, 1, \dodoi{https://doi.org/10.1029/JA077i001p00001}

\bibitem[{Cranmer(2020)}]{Cranmer_2020}
Cranmer, S.~R. 2020, Research Notes of the {AAS}, 4, 249,
  \dodoi{10.3847/2515-5172/abd5ae}

\bibitem[{Feldman {et~al.}(1975)Feldman, Asbridge, Bame, Montgomery, \&
  Gary}]{https://doi.org/10.1029/JA080i031p04181}
Feldman, W.~C., Asbridge, J.~R., Bame, S.~J., Montgomery, M.~D., \& Gary, S.~P.
  1975, Journal of Geophysical Research (1896-1977), 80, 4181,
  \dodoi{https://doi.org/10.1029/JA080i031p04181}

\bibitem[{Gurnett \& Bhattacharjee(2017)}]{gurnett_bhattacharjee_2017}
Gurnett, D.~A., \& Bhattacharjee, A. 2017, Introduction to Plasma Physics: With
  Space, Laboratory and Astrophysical Applications, 2nd edn. (Cambridge
  University Press), \dodoi{10.1017/9781139226059}

\bibitem[{Halekas {et~al.}(2021)Halekas, {Whittlesey, P. L.}, {Larson, D. E.},
  {McGinnis, D.}, {Bale, S. D.}, {Berthomier, M.}, {Case, A. W.}, {Chandran, B.
  D. G.}, {Kasper, J. C.}, {Klein, K. G.}, {Korreck, K. E.}, {Livi, R.},
  {MacDowall, R. J.}, {Maksimovic, M.}, {Malaspina, D. M.}, {Matteini, L.},
  {Pulupa, M. P.}, \& {Stevens, M. L.}}]{refId00}
Halekas, J.~S., {Whittlesey, P. L.}, {Larson, D. E.}, {et~al.} 2021, A\&A, 650,
  A15, \dodoi{10.1051/0004-6361/202039256}

\bibitem[{{Horaites} {et~al.}(2018{\natexlab{a}}){Horaites}, {Astfalk},
  {Boldyrev}, \& {Jenko}}]{2018MNRAS.480.1499H}
{Horaites}, K., {Astfalk}, P., {Boldyrev}, S., \& {Jenko}, F.
  2018{\natexlab{a}}, \mnras, 480, 1499, \dodoi{10.1093/mnras/sty1808}

\bibitem[{{Horaites} {et~al.}(2019){Horaites}, {Boldyrev}, \&
  {Medvedev}}]{2019MNRAS.484.2474H}
{Horaites}, K., {Boldyrev}, S., \& {Medvedev}, M.~V. 2019, \mnras, 484, 2474,
  \dodoi{10.1093/mnras/sty3504}

\bibitem[{{Horaites} {et~al.}(2018{\natexlab{b}}){Horaites}, {Boldyrev},
  {Wilson}, {Vi{\~n}as}, \& {Merka}}]{2018MNRAS.474..115H}
{Horaites}, K., {Boldyrev}, S., {Wilson}, Lynn~B., I., {Vi{\~n}as}, A.~F., \&
  {Merka}, J. 2018{\natexlab{b}}, \mnras, 474, 115,
  \dodoi{10.1093/mnras/stx2555}

\bibitem[{Isenberg(1997)}]{https://doi.org/10.1029/96JA03671}
Isenberg, P.~A. 1997, Journal of Geophysical Research: Space Physics, 102,
  4719, \dodoi{https://doi.org/10.1029/96JA03671}

\bibitem[{Jeong {et~al.}(2020)Jeong, Verscharen, Wicks, \&
  Fazakerley}]{Jeong_2020}
Jeong, S.-Y., Verscharen, D., Wicks, R.~T., \& Fazakerley, A.~N. 2020, The
  Astrophysical Journal, 902, 128, \dodoi{10.3847/1538-4357/abb099}

\bibitem[{{Jockers}(1970)}]{1970A&A.....6..219J}
{Jockers}, K. 1970, \aap, 6, 219

\bibitem[{{Kasper} {et~al.}(2016){Kasper}, {Abiad}, {Austin}, {Balat-Pichelin},
  {Bale}, {Belcher}, {Berg}, {Bergner}, {Berthomier}, {Bookbinder}, {Brodu},
  {Caldwell}, {Case}, {Chandran}, {Cheimets}, {Cirtain}, {Cranmer}, {Curtis},
  {Daigneau}, {Dalton}, {Dasgupta}, {DeTomaso}, {Diaz-Aguado}, {Djordjevic},
  {Donaskowski}, {Effinger}, {Florinski}, {Fox}, {Freeman}, {Gallagher},
  {Gary}, {Gauron}, {Gates}, {Goldstein}, {Golub}, {Gordon}, {Gurnee}, {Guth},
  {Halekas}, {Hatch}, {Heerikuisen}, {Ho}, {Hu}, {Johnson}, {Jordan},
  {Korreck}, {Larson}, {Lazarus}, {Li}, {Livi}, {Ludlam}, {Maksimovic},
  {McFadden}, {Marchant}, {Maruca}, {McComas}, {Messina}, {Mercer}, {Park},
  {Peddie}, {Pogorelov}, {Reinhart}, {Richardson}, {Robinson}, {Rosen},
  {Skoug}, {Slagle}, {Steinberg}, {Stevens}, {Szabo}, {Taylor}, {Tiu}, {Turin},
  {Velli}, {Webb}, {Whittlesey}, {Wright}, {Wu}, \&
  {Zank}}]{2016SSRv..204..131K}
{Kasper}, J.~C., {Abiad}, R., {Austin}, G., {et~al.} 2016, \ssr, 204, 131,
  \dodoi{10.1007/s11214-015-0206-3}

\bibitem[{Kolobov {et~al.}(2020)Kolobov, Arslanbekov, \& Levko}]{2020}
Kolobov, V., Arslanbekov, R., \& Levko, D. 2020, 1623, 012006,
  \dodoi{10.1088/1742-6596/1623/1/012006}

\bibitem[{Landi {et~al.}(2012)Landi, Matteini, \& Pantellini}]{Landi_2012}
Landi, S., Matteini, L., \& Pantellini, F. 2012, The Astrophysical Journal,
  760, 143, \dodoi{10.1088/0004-637x/760/2/143}

\bibitem[{Landi \& Pantellini(2003)}]{refId0}
Landi, S., \& Pantellini, F. 2003, A\&A, 400, 769,
  \dodoi{10.1051/0004-6361:20021822}

\bibitem[{le~Roux \& Webb(2009)}]{le_Roux_2009}
le~Roux, J.~A., \& Webb, G.~M. 2009, The Astrophysical Journal, 693, 534,
  \dodoi{10.1088/0004-637x/693/1/534}

\bibitem[{{le Roux} {et~al.}(2007){le Roux}, {Webb}, {Florinski}, \&
  {Zank}}]{2007ApJ...662..350L}
{le Roux}, J.~A., {Webb}, G.~M., {Florinski}, V., \& {Zank}, G.~P. 2007, \apj,
  662, 350, \dodoi{10.1086/517601}

\bibitem[{{Lie-Svendsen} {et~al.}(1997){Lie-Svendsen}, {Hansteen}, \&
  {Leer}}]{1997JGR...102.4701L}
{Lie-Svendsen}, {\O}., {Hansteen}, V.~H., \& {Leer}, E. 1997, \jgr, 102, 4701,
  \dodoi{10.1029/96JA03632}

\bibitem[{{Lie-Svendsen} \& {Leer}(2000)}]{2000JGR...105...35L}
{Lie-Svendsen}, {\O}., \& {Leer}, E. 2000, \jgr, 105, 35,
  \dodoi{10.1029/1999JA900438}

\bibitem[{{Lindquist}(1966)}]{1966AnPhy..37..487L}
{Lindquist}, R.~W. 1966, Annals of Physics, 37, 487,
  \dodoi{10.1016/0003-4916(66)90207-7}

\bibitem[{Livadiotis \& McComas(2009)}]{https://doi.org/10.1029/2009JA014352}
Livadiotis, G., \& McComas, D.~J. 2009, Journal of Geophysical Research: Space
  Physics, 114, \dodoi{https://doi.org/10.1029/2009JA014352}

\bibitem[{{Livadiotis} \& {McComas}(2013)}]{2013SSRv..175..183L}
{Livadiotis}, G., \& {McComas}, D.~J. 2013, \ssr, 175, 183,
  \dodoi{10.1007/s11214-013-9982-9}

\bibitem[{Livi {et~al.}(1986)Livi, Marsch, \&
  Rosenbauer}]{https://doi.org/10.1029/JA091iA07p08045}
Livi, S., Marsch, E., \& Rosenbauer, H. 1986, Journal of Geophysical Research:
  Space Physics, 91, 8045, \dodoi{https://doi.org/10.1029/JA091iA07p08045}

\bibitem[{Ljepojevic {et~al.}(1990)Ljepojevic, Burgess, \&
  Moffatt}]{doi:10.1098/rspa.1990.0026}
Ljepojevic, N.~N., Burgess, A., \& Moffatt, H.~K. 1990, Proceedings of the
  Royal Society of London. A. Mathematical and Physical Sciences, 428, 71,
  \dodoi{10.1098/rspa.1990.0026}

\bibitem[{L{\'{o}}pez {et~al.}(2020)L{\'{o}}pez, Lazar, Shaaban, Poedts, \&
  Moya}]{L_pez_2020}
L{\'{o}}pez, R.~A., Lazar, M., Shaaban, S.~M., Poedts, S., \& Moya, P.~S. 2020,
  The Astrophysical Journal, 900, L25, \dodoi{10.3847/2041-8213/abaf56}

\bibitem[{{Maksimovic} {et~al.}(1997){Maksimovic}, {Pierrard}, \&
  {Lemaire}}]{1997A&A...324..725M}
{Maksimovic}, M., {Pierrard}, V., \& {Lemaire}, J.~F. 1997, \aap, 324, 725

\bibitem[{Maksimovic {et~al.}(2005)Maksimovic, Zouganelis, Chaufray, Issautier,
  Scime, Littleton, Marsch, McComas, Salem, Lin, \&
  Elliott}]{https://doi.org/10.1029/2005JA011119}
Maksimovic, M., Zouganelis, I., Chaufray, J.-Y., {et~al.} 2005, Journal of
  Geophysical Research: Space Physics, 110,
  \dodoi{https://doi.org/10.1029/2005JA011119}

\bibitem[{{Marsch}(2006)}]{2006LRSP....3....1M}
{Marsch}, E. 2006, Living Reviews in Solar Physics, 3, 1,
  \dodoi{10.12942/lrsp-2006-1}

\bibitem[{{Marsch} {et~al.}(1989){Marsch}, {Pilipp}, {Thieme}, \&
  {Rosenbauer}}]{1989JGR....94.6893M}
{Marsch}, E., {Pilipp}, W.~G., {Thieme}, K.~M., \& {Rosenbauer}, H. 1989, \jgr,
  94, 6893, \dodoi{10.1029/JA094iA06p06893}

\bibitem[{Micera {et~al.}(2021)Micera, Zhukov, L{\'{o}}pez, Boella, Tenerani,
  Velli, Lapenta, \& Innocenti}]{Micera_2021}
Micera, A., Zhukov, A.~N., L{\'{o}}pez, R.~A., {et~al.} 2021, The Astrophysical
  Journal, 919, 42, \dodoi{10.3847/1538-4357/ac1067}

\bibitem[{Micera {et~al.}(2020)Micera, Zhukov, L{\'{o}}pez, Innocenti, Lazar,
  Boella, \& Lapenta}]{Micera_2020}
---. 2020, The Astrophysical Journal, 903, L23,
  \dodoi{10.3847/2041-8213/abc0e8}

\bibitem[{Moncuquet {et~al.}(2020)Moncuquet, Meyer-Vernet, Issautier, Pulupa,
  Bonnell, Bale, de~Wit, Goetz, Griton, Harvey, MacDowall, Maksimovic, \&
  Malaspina}]{Moncuquet_2020}
Moncuquet, M., Meyer-Vernet, N., Issautier, K., {et~al.} 2020, The
  Astrophysical Journal Supplement Series, 246, 44,
  \dodoi{10.3847/1538-4365/ab5a84}

\bibitem[{{Nicolaou} \& {Livadiotis}(2016)}]{2016Ap&SS.361..359N}
{Nicolaou}, G., \& {Livadiotis}, G. 2016, \apss, 361, 359,
  \dodoi{10.1007/s10509-016-2949-z}

\bibitem[{Nolting(2016)}]{Nolting2016}
Nolting, W. 2016, Hamilton Mechanics (Cham: Springer International Publishing),
  101--173, \dodoi{10.1007/978-3-319-40129-4_2}

\bibitem[{Owens {et~al.}(2008)Owens, Crooker, \&
  Schwadron}]{https://doi.org/10.1029/2008JA013294}
Owens, M.~J., Crooker, N.~U., \& Schwadron, N.~A. 2008, Journal of Geophysical
  Research: Space Physics, 113, \dodoi{https://doi.org/10.1029/2008JA013294}

\bibitem[{{Parker}(1958)}]{1958ApJ...128..664P}
{Parker}, E.~N. 1958, \apj, 128, 664, \dodoi{10.1086/146579}

\bibitem[{Pierrard {et~al.}(2001)Pierrard, Issautier, Meyer-Vernet, \&
  Lemaire}]{https://doi.org/10.1029/2000GL011888}
Pierrard, V., Issautier, K., Meyer-Vernet, N., \& Lemaire, J. 2001, Geophysical
  Research Letters, 28, 223, \dodoi{https://doi.org/10.1029/2000GL011888}

\bibitem[{Pilipp {et~al.}(1987{\natexlab{a}})Pilipp, Miggenrieder, Montgomery,
  Mühlhäuser, Rosenbauer, \& Schwenn}]{doi:10.1029/JA092iA02p01075}
Pilipp, W.~G., Miggenrieder, H., Montgomery, M.~D., {et~al.}
  1987{\natexlab{a}}, Journal of Geophysical Research: Space Physics, 92, 1075,
  \dodoi{10.1029/JA092iA02p01075}

\bibitem[{Pilipp {et~al.}(1987{\natexlab{b}})Pilipp, Miggenrieder,
  Mühlhäuser, Rosenbauer, Schwenn, \&
  Neubauer}]{https://doi.org/10.1029/JA092iA02p01103}
Pilipp, W.~G., Miggenrieder, H., Mühlhäuser, K.~H., {et~al.}
  1987{\natexlab{b}}, Journal of Geophysical Research: Space Physics, 92, 1103,
  \dodoi{https://doi.org/10.1029/JA092iA02p01103}

\bibitem[{Rosenbluth {et~al.}(1957)Rosenbluth, MacDonald, \&
  Judd}]{PhysRev.107.1}
Rosenbluth, M.~N., MacDonald, W.~M., \& Judd, D.~L. 1957, Phys. Rev., 107, 1,
  \dodoi{10.1103/PhysRev.107.1}

\bibitem[{{Schroeder} {et~al.}(2021){Schroeder}, {Boldyrev}, \&
  {Astfalk}}]{2021MNRAS.507.1329S}
{Schroeder}, J.~M., {Boldyrev}, S., \& {Astfalk}, P. 2021, \mnras, 507, 1329,
  \dodoi{10.1093/mnras/stab2228}

\bibitem[{{Scudder}(1992{\natexlab{a}})}]{1992ApJ...398..319S}
{Scudder}, J.~D. 1992{\natexlab{a}}, \apj, 398, 319, \dodoi{10.1086/171859}

\bibitem[{{Scudder}(1992{\natexlab{b}})}]{1992ApJ...398..299S}
---. 1992{\natexlab{b}}, \apj, 398, 299, \dodoi{10.1086/171858}

\bibitem[{{Scudder}(2019)}]{2019ApJ...882..146S}
---. 2019, \apj, 882, 146, \dodoi{10.3847/1538-4357/ab3348}

\bibitem[{Seough {et~al.}(2015)Seough, Nariyuki, Yoon, \& Saito}]{Seough_2015}
Seough, J., Nariyuki, Y., Yoon, P.~H., \& Saito, S. 2015, The Astrophysical
  Journal, 811, L7, \dodoi{10.1088/2041-8205/811/1/l7}

\bibitem[{{Skilling}(1971)}]{1971ApJ...170..265S}
{Skilling}, J. 1971, \apj, 170, 265, \dodoi{10.1086/151210}

\bibitem[{Smith {et~al.}(2012)Smith, Marsch, \& Helander}]{Smith_2012}
Smith, H.~M., Marsch, E., \& Helander, P. 2012, The Astrophysical Journal, 753,
  31, \dodoi{10.1088/0004-637x/753/1/31}

\bibitem[{{\v{S}tver{\'a}k} {et~al.}(2009){\v{S}tver{\'a}k}, Maksimovic,
  Tr\'avn\'i{\v c}ek, Marsch, Fazakerley, \& Scime}]{doi:10.1029/2008JA013883}
{\v{S}tver{\'a}k}, {\v S}., Maksimovic, M., Tr\'avn\'i{\v c}ek, P.~M., {et~al.}
  2009, Journal of Geophysical Research: Space Physics, 114,
  \dodoi{10.1029/2008JA013883}

\bibitem[{{\v{S}tver{\'a}k} {et~al.}(2015){\v{S}tver{\'a}k}, Trávníček, \&
  Hellinger}]{https://doi.org/10.1002/2015JA021368}
{\v{S}tver{\'a}k}, {\v S}., Trávníček, P.~M., \& Hellinger, P. 2015, Journal
  of Geophysical Research: Space Physics, 120, 8177,
  \dodoi{https://doi.org/10.1002/2015JA021368}

\bibitem[{Sun {et~al.}(2021)Sun, Zhao, Liu, Voitenko, Pierrard, Shi, Yao, Xie,
  \& Wu}]{Sun_2021}
Sun, H., Zhao, J., Liu, W., {et~al.} 2021, The Astrophysical Journal Letters,
  916, L4, \dodoi{10.3847/2041-8213/ac0f02}

\bibitem[{Tang {et~al.}(2020)Tang, Zank, \& Kolobov}]{Tang_2020}
Tang, B., Zank, G.~P., \& Kolobov, V.~I. 2020, The Astrophysical Journal, 892,
  95, \dodoi{10.3847/1538-4357/ab7a93}

\bibitem[{{Vasko} {et~al.}(2019){Vasko}, {Krasnoselskikh}, {Tong}, {Bale},
  {Bonnell}, \& {Mozer}}]{2019ApJ...871L..29V}
{Vasko}, I.~Y., {Krasnoselskikh}, V., {Tong}, Y., {et~al.} 2019, \apjl, 871,
  L29, \dodoi{10.3847/2041-8213/ab01bd}

\bibitem[{{Verscharen} {et~al.}(2019){Verscharen}, {Chandran}, {Jeong},
  {Salem}, {Pulupa}, \& {Bale}}]{2019ApJ...886..136V}
{Verscharen}, D., {Chandran}, B. D.~G., {Jeong}, S.-Y., {et~al.} 2019, \apj,
  886, 136, \dodoi{10.3847/1538-4357/ab4c30}

\bibitem[{Verscharen {et~al.}(2019)Verscharen, Klein, \&
  Maruca}]{d437a059c4a9467cbb87b86c0f01b676}
Verscharen, D., Klein, K., \& Maruca, B. 2019, Living Reviews in Solar Physics,
  16, \dodoi{10.1007/s41116-019-0021-0}

\bibitem[{{Vi{\~n}as} {et~al.}(2000){Vi{\~n}as}, {Wong}, \&
  {Klimas}}]{2000ApJ...528..509V}
{Vi{\~n}as}, A.~F., {Wong}, H.~K., \& {Klimas}, A.~J. 2000, \apj, 528, 509,
  \dodoi{10.1086/308151}

\bibitem[{Vocks(2002)}]{Vocks_2002}
Vocks, C. 2002, The Astrophysical Journal, 568, 1017, \dodoi{10.1086/338884}

\bibitem[{Vocks \& Mann(2003)}]{Vocks_2003}
Vocks, C., \& Mann, G. 2003, The Astrophysical Journal, 593, 1134,
  \dodoi{10.1086/376682}

\bibitem[{{Webb}(1985)}]{1985ApJ...296..319W}
{Webb}, G.~M. 1985, \apj, 296, 319, \dodoi{10.1086/163451}

\bibitem[{{Whittlesey} {et~al.}(2020){Whittlesey}, {Larson}, {Kasper},
  {Halekas}, {Abatcha}, {Abiad}, {Berthomier}, {Case}, {Chen}, {Curtis},
  {Dalton}, {Klein}, {Korreck}, {Livi}, {Ludlam}, {Marckwordt}, {Rahmati},
  {Robinson}, {Slagle}, {Stevens}, {Tiu}, \& {Verniero}}]{2020ApJS..246...74W}
{Whittlesey}, P.~L., {Larson}, D.~E., {Kasper}, J.~C., {et~al.} 2020, \apjs,
  246, 74, \dodoi{10.3847/1538-4365/ab7370}

\bibitem[{Yakovlev \& Pisanko(2018)}]{YAKOVLEV2018552}
Yakovlev, O.~I., \& Pisanko, Y.~V. 2018, Advances in Space Research, 61, 552,
  \dodoi{https://doi.org/10.1016/j.asr.2017.10.052}

\bibitem[{Zank(2013)}]{book}
Zank, G. 2013, Transport Processes in Space Physics and Astrophysics, Vol. 877,
  \dodoi{10.1007/978-1-4614-8480-6}

\bibitem[{Zouganelis {et~al.}(2005)Zouganelis, Meyer-Vernet, Landi, Maksimovic,
  \& Pantellini}]{Zouganelis_2005}
Zouganelis, I., Meyer-Vernet, N., Landi, S., Maksimovic, M., \& Pantellini, F.
  2005, The Astrophysical Journal, 626, L117, \dodoi{10.1086/431904}

\end{thebibliography}
\bibliographystyle{aasjournal}



\end{document}